\documentclass[a4paper]{article}
\usepackage{amsmath, amsfonts}

\usepackage{amssymb}

\makeatletter

\AtBeginDocument{\DeclareFontEncoding{T2A}{}{}}

\makeatother

\newtheorem{theorem}{Theorem}[section]

\newtheorem{lemma}{Lemma}[section]

\newtheorem{remark}{Remark}[section]

\let\epsilon=\varepsilon
\let\phi=\varphi
\def\tire{\thinspace--\thinspace}
\def\<{\langle}
\def\>{\rangle}

\providecommand{\keywords}[1]
{
  \small	
  \textbf{\textit{Keywords:}} #1
}

\title{Introduction to Micro Life of Graphs. I.}
\author{V.A.~Malyshev$^{1}$,   A.A.~Zamyatin$^{1}$ \\
\small 
\textit{$^{1}$Lomonosov Moscow State University} \\
}
\date{}

\begin{document}
\allowdisplaybreaks

\maketitle



\begin{abstract}
 Multiparticle systems on complicated metric graphs might have many
applications in physics, biology and social life. But the corresponding
science still does not exist. Here we start it with simplest examples
where there is quadratic interaction between neighboring particles
and deterministic external forces. In this introduction we consider
stable configurations and stable flows on one and two edge graphs.
Moreover, distribution of mean (as in virial theorem) kinetic and
potential energies along the graph is considered.  
  \end{abstract}


  \vspace{-5pt}
  
\tableofcontents

\section{Introduction}

Mathematician, who besides pure mathematical problems, wants to do
something else, might ask the question -- could mathematics give some
ideas why our life is so short.

More exact question could be like this: can mathematics provide some
kind of broad view on bio organism using small number of axioms at
the upper level but many examples on the lower. 

Immediately mathematician sees that the amount of information concerning
physics, chemistry and biology of bio organisms is already immense
but is still growing faster and faster. Moreover, even the existence
of long molecules and solids does not have rigorous proofs neither
in classical (assuming only Newton equation with Coulomb force), nor
in quantum mathematical physics. This suggests the answer to this
question -- obviously NOT. 

However, some of us can remember that ``toy models'' were popular
in mathematical statistical physics in second half of last century.
So, we could try to find toy models of our own health, in particular
due to harmful external influence. 

The components of such toy models are the following:

1) we consider large systems of point particles such that their micro
behavior gives rise to macro effects which everyone knows when feels
his own body;

2) this system, as a whole, can be imagined as ``body'', and edges
(or subsets of edges) are the parts of this body. For sufficiently
large graphs one can imagine even more complicated hierarchy; 

3) for such system there are different static and dynamical problems.
For example, stable state (no dynamics) is the minimum of the potential
energy. It is important to understand what (potential and kinetic)
energy distribution can be over living ``body''. If the initial
conditions are not stable, or if external forces are time-dependent,
then potential and kinetic energy can be quite differently distributed
over the ``body''. It is obvious that any part of the body should
have sufficient energy to survive;

4) if our graph has cycles, then stationary flows (like electric current
for Coulomb forces) along some or all cycles are possible; 

5) in the first model the particles move in their local potential
wells which are formed by neighboring particles. Next development
of the model is to introduce external media and flows in this media.
Moreover, this interaction allows, for any edge, departure and arrival
of particles. Departure can be in cases when dynamical situation leads
to collisions or too close rapprochement of neighboring particles.
On the contrary, arrival can occur if neighboring particles on the
edge become too far apart from each other;

6) growth of graph becomes also possible if we consider the edge not
as the segment of fixed length, but of the length which is defined
as the sum of distances between neighboring edges. Then the breaks
and growth of edges and even appearance of new edges is possible. 

Now we start rigorous definitions.

\paragraph{Static metric graph}

In this case we consider many-particle systems on metric graphs, where
(static) metric graph $G$ is a graph with metrics, where each edge
is metrically isomorphic to a segment of the real line, and the distance
between two points is the minimal length of path between these points. 

\smallskip

On each edge $l$ there is large system of identical particles. Potential
energy of such system is the sum
\[
U=\sum_{(i,j)}u(\rho_{ij})
\]
over pairs $i,j$ of particles, for which there exists a path between
$i$ and $j$ such that does not pass over any other particle. And
$\rho_{ij}$ is the minimal length of such path. In this paper we
start with the case when the interaction is quadratic, that is 
\[
u(\rho_{ij})=\frac{\omega_{ij}^{2}}{2}\rho_{ij}^{2}.
\]
Moreover, external forces can act on some or all particles.

\paragraph{Dynamic graph}

Here abstract graph $G$ is fixed. We assume it connected and not
more than one edge between any two vertices. Vertices of graph may
be called point particles and numerated as $i=1,\ldots ,N$. Existence
of edge $l=(i.j)$ between vertices means that these two ``particles''
interact. Each edge has variable length $q_{l}=q_{ij}>0$. Dynamics
of these lengths is defined by Newton equations
\[
\frac{d^{2}q_{l}}{dt^{2}}=F_{l}(q_{l})+\sum_{m:m\neq l}F_{lm}(q_{l},q_{m}),
\]
where the summation is over all edges $m$ such that have common vertex
with edge $l$. And $F_{l},F_{lm}$ are forces -- real functions of
one and two variables correspondingly. 

\smallskip

In this introduction we consider only static graphs and give an introduction
to the statics of such systems for three simplest graphs -- segment,
circle and graph with 2 edges. The first interesting topics -- ground
states of such systems. 

\smallskip

Ground states with different interactions, but without external force,
for classical finite and infinite particle systems were intensively
studied during last 30 years, see for example [1--6]. Main results
in these papers concern periodicity of the ground states. In the papers
[7--14] mostly Coulomb systems were considered.

\smallskip

Other question in our paper: 1) flow of particles along the cycle;
2) energy distribution on parts of the graph. 

\section{Stationary flow of point particles}

\paragraph{Periodic equilibrium configurations}

Let $S=S_{L}$ be circle of length $L$, or segment $[0,L]$ with
identified end points. Consider $N$ point particles $0,1,\ldots ,$ $N-1$
with coordinates 
\begin{equation}
0=x_{0}<x_{1}<\ldots <x_{N-1}<L\label{increasing_conf}
\end{equation}
 or infinite periodic sequence (with period $L$) on the real axis
$R$ 
\[
\ldots <x_{-1}<0=x_{0}<x_{1}<\ldots <x_{N-1}<x_{N}=L<x_{N+1}<\ldots 
\]
where $x_{k+N}=x_{k}+L$ for any $k$. We assume formal potential
energy 
\[
U=\frac{\omega^{2}}{2}\sum_{k\in Z}(x_{k}-x_{k-1}-a)^{2}
\]
with $a=L/N$, and moreover there are the following external
forces:

1. constant force $f>0$ on the particle $0$, or on any particle
$iN,i\in Z,$;

2. constant forces $-\phi<0$ (that is $\phi>0$) on any particle.

Define $\epsilon_{k}$ as
\[
\Delta_{k}=x_{k}-x_{k-1}=\frac{L}{N}+\frac{\epsilon_{k}}{N^{2}}.
\]
Then: the sequence $\epsilon_{k}$ is also periodic, $-LN<\epsilon_{k}$
for any $k$, and
\begin{equation}
\sum_{k=1}^{N}\epsilon_{k}=0. \label{sum_epsilon}
\end{equation}
Configuration is called equilibrium, if the force, acting on any particle,
is zero. 

\begin{theorem}
There exists fixed (equilibrium) configuration, satisfying condition
(\ref{increasing_conf}), iff the following two conditions hold:
\begin{itemize}
\item[{\rm 1)}]
$\phi=f/N,$
\item[{\rm 2)}] 
$\displaystyle 
\frac{f}{2\omega^{2}L}<\frac{1}{N(1-1/N)}.$
\end{itemize}

This configuration is unique and is defined by
\[
\epsilon_{1}=-\frac{fN(N-1)}{2\omega^{2}}, \quad \epsilon_{N}=\frac{fN(N-1)}{2\omega^{2}},
\]
\[
\epsilon_{k}=\epsilon_{1}+(k-1)\frac{fN}{\omega^{2}},\quad k=1,\ldots ,N-1.
\]

If moreover
\[
\frac{f}{2\omega^{2}L}<\frac{C}{N^{2}}
\]
for some fixed constant $C>0$, then, in such scaling, $\epsilon_{k}=O(1)$
uniformly in $k=1,\ldots ,N$. 
\end{theorem}

\medskip\noindent
{\it Proof}. The condition that the total force on particles $k=1,\ldots ,N-1$
is zero can be written as follows
\begin{equation}
\omega^{2}\Bigl(\frac{\epsilon_{k+1}}{N^{2}}-\frac{\epsilon_{k}}{N^{2}}\Bigr)=\phi\Longleftrightarrow\epsilon_{k+1}-\epsilon_{k}=\frac{\phi N^{2}}{\omega^{2}}\label{except_0}
\end{equation}
and on the particle $0$ (or $N$)
\begin{equation}
  f+\omega^{2}\Bigl(\frac{\epsilon_{1}}{N^{2}}-\frac{\epsilon_{N}}{N^{2}}\Bigr)=\phi\Longleftrightarrow\epsilon_{1}-\epsilon_{N}=\frac{N^{2}}{\omega^{2}}(-f+\phi).
  \label{0_particle}
\end{equation}
Summation of (\ref{0_particle}) and all (\ref{except_0}) gives
\begin{equation}
(N-1)\phi-f+\phi=0\Longleftrightarrow\phi=\frac{f}{N}. \label{phi}
\end{equation}
Thus, condition 1) of the Theorem is a necessary condition. 

And we can rewrite conditions (\ref{except_0}) and (\ref{0_particle})
as
\[
\epsilon_{k+1}-\epsilon_{k}=\frac{fN}{\omega^{2}},\quad \epsilon_{1}-\epsilon_{N}=-\frac{fN(N-1)}{\omega^{2}}.
\]
Then 
\[
\epsilon_{k}=\epsilon_{1}+(k-1)\frac{fN}{\omega^{2}}, \quad k=2,\ldots ,N-1,\quad \epsilon_{N}=\epsilon_{1}+\frac{fN(N-1)}{\omega^{2}}.
\]
Substituting to (\ref{sum_epsilon}) we get 
\[
N\epsilon_{1}+\frac{fN}{\omega^{2}}\sum_{k=2}^{N-1}(k-1)+\frac{fN(N-1)}{\omega^{2}}=N\epsilon_{1}+\frac{fN}{\omega^{2}}\sum_{k=2}^{N}(k-1)=
\]
\[
=N(\epsilon_{1}+\frac{fN(N-1)}{2\omega^{2}})=0
\]
or
\[
\epsilon_{1}=-\frac{fN(N-1)}{2\omega^{2}}
\]
and then
\[
\epsilon_{N}=\frac{fN(N-1)}{2\omega^{2}}.
\]
Also for all $k=1,\ldots ,N-1$
\[
\epsilon_{k}=\epsilon_{1}+(k-1)\frac{fN}{\omega^{2}}.
\]
Note that 
\[
x_{k}>x_{k-1}\Longleftrightarrow\frac{\epsilon_{k}}{N^{2}}>-\frac{L}{N}\Longleftrightarrow\frac{\epsilon_{1}}{N^{2}}=-\frac{fN(N-1)}{2\omega^{2}N^{2}}>-\frac{L}{N}\Longleftrightarrow
\]
\begin{equation}
\Longleftrightarrow\frac{f}{2\omega^{2}}<\frac{L}{N(1-1/N)}\Longleftrightarrow\frac{f}{2\omega^{2}L}<\frac{1}{N(1-1/N)}. \label{scaling}
\end{equation}
That is for fixed $L$, the parameters $f,\omega$ should be scaled
(with respect to $N$) so that $f/(2\omega^{2}L)< 1/N$
hold.

\paragraph{Stationary flow}

Now we want to prove that there exists stationary periodic flow of
particles as
\[
x_{k}(t)=x_{k}(0)+vt,
\]
where $v>0$ and $x_{k}(0)=x_{k}$, defined in the Theorem. That is
the particles move with the same velocity $v>0$. This flow is driven
by the constant force $f>0$ acting only on the particle $0$ and
by some dissipative force which acts on any particle and depends only
on its velocity. Such dissipative forces $-g(v)$, where $g(v)$ is
positive smooth increasing function) (often used is $g(v)=\alpha v$
with some $\alpha>0$). Assume that we fixed this function $g$. Then
there exists unique $v>0$ such that $g(v)=\phi=f/N$. 

This construction is resembles the famous Drude's model of electric
current (one can find it in any text book on electricity) where particles
move without interaction under the influence of external constant
force and dissipative force, acting on all particles. Here however
the particles move due to the driven force acting on only one particles.

Even more realistic model with Coulomb interaction forces see in \cite{Mal_Self}.

\section{Static equilibrium}

Here we give 5 examples of 1-dimensional stable equilibrium configurations. 

\paragraph{External force on one particle}

Consider $N+1$ particles $i=0,1,\ldots ,N$ on $R$ with coordinates
$x_{i}$. Potential energy is assumed to be 
\begin{equation}
U=-fx_{0}+\frac{\omega_{0}^{2}}{2}x_{0}^{2}+\frac{\omega_{1}^{2}}{2}\sum_{i=1}^{N}(x_{i}-x_{i-1}-a)^{2}. \label{U_2}
\end{equation}
That is there is constant force $f$ which acts only on particle $0$.
Moreover, particle $N$ is assumed to be tightly fixed, that is $x_{N}\equiv L>0$.
We see from (\ref{U_2}) that particle $0$ is attached to $0\in R$
by harmonic force. 

\begin{theorem}\label{th0}
For any parameters $f,L>0,\omega_{0},\omega_{1}>0$ equilibrium configuration
exists and is unique. Two cases are possible:

1) If $f<\omega_{0}^{2}L+\omega_{1}^{2}a$ then 
\[
x_{0}=\frac{\omega_{1}^{2}(L-Na)+Nf}{\omega_{1}^{2}+\omega_{0}^{2}N},
\]
\[
x_{k}=x_{0}+\frac{\omega_{0}^{2}L+\omega_{1}^{2}a-f}{\omega_{1}^{2}+\omega_{0}^{2}N}k,\quad k=1,\ldots ,N-1.
\]
It follows that $x_{0}<x_{1}<\dots<x_{N}=L$.

2) If $f\geq\omega_{0}^{2}L+\omega_{1}^{2}a$ then $x_{0}=x_{1}=\dots=x_{N}=L$.

In all cases this equilibrium is stable, that is the minimum of
$U$.
\end{theorem}

Note that a ``natural order'' $0\leq x_{0}<x_{1}<\dots<x_{N}=L$ holds
iff
\[
\omega_{1}^{2}\Bigl(a-\frac{L}{N}\Bigr)\leq f<\omega_{0}^{2}L+\omega_{1}^{2}a.
\]

Now consider the case when $x_{N}$ is not fixed but the potential
energy is
\[
U=-fx_{0}+\frac{\omega_{0}^{2}}{2}x_{0}^{2}+\frac{\omega_{0}^{2}}{2}(x_{N}-L)^{2}+\frac{\omega_{1}^{2}}{2}\sum_{i=1}^{N}(x_{i}-x_{i-1}-a)^{2}.
\]

\begin{theorem} \label{th1}
$\phantom{ai}$ For any parameters equilibrium configuration exists and is unique.
It is given by
\[
x_{0}=\frac{\omega_{1}^{2}\left(L+N( f / \omega_{1}^{2}-a)\right)+\omega_{1}^{2}\omega_{0}^{-2}f}{2\omega_{1}^{2}+N\omega_{0}^{2}},
\]
\[
x_{k}=x_{0}+\frac{\omega_{0}^{2}L+2\omega_{1}^{2}a-f}{2\omega_{1}^{2}+N\omega_{0}^{2}}k,\;\; k=1,\ldots ,N.
\]
There can be 3 types of equilibrium configurations:
\[
x_{0}<x_{1}<\dots<x_{N}\Longleftrightarrow f<\omega_{0}^{2}L+2\omega_{1}^{2}a
\]
\[
x_{0}=x_{1}=\dots=x_{N}=L+\frac{\omega_{1}^{2}a(N+2\omega_{1}^{2}\omega_{0}^{-2})}{2\omega_{1}^{2}+N\omega_{0}^{2}}\Longleftrightarrow f=\omega_{0}^{2}L+2\omega_{1}^{2}a
\]
\[
x_{0}>x_{1}>\dots>x_{N}>L\Longleftrightarrow f>\omega_{0}^{2}L+2\omega_{1}^{2}a
\]
\end{theorem}

Note that $0\leq x_{0}<x_{1}<\dots<x_{N}\leq L$ iff $L\geq aN$ and
\[
-\frac{\omega_{1}^{2}(L-aN)}{N+\omega_{1}^{2}\omega_{0}^{-2}}\leq f\leq\omega_{0}^{2}(L-aN).
\]

\paragraph{Forces on both extreme points}

Here we assume that force $f$ acts on particle $0,$ and the force
$-f$ acts on the particle $N.$ That is the potential energy is
\[
U=\frac{\omega_{0}^{2}}{2}x_{0}^{2}+\frac{\omega_{0}^{2}}{2}(x_{N}-L)^{2}+\frac{\omega_{1}^{2}}{2}\sum_{i=1}^{N}(x_{i}-x_{i-1}-a)^{2}-fx_{0}+fx_{N}.
\]

\begin{theorem}\label{th2}
  $\phantom{a}$
  \begin{itemize}
\item[{\rm 1)}] For any parameters equilibrium configuration exists, is unique
and the coordinates are given by
\[
x_{0}=\frac{\omega_{1}^{2}\left(L+N( f / \omega_{1}^{2} -a)\right)}{2\omega_{1}^{2}+N\omega_{0}^{2}},
\]
\[
x_{k}-x_{k-1}=\frac{\omega_{0}^{2}L+2\omega_{1}^{2}a-2f}{2\omega_{1}^{2}+N\omega_{0}^{2}},\quad k=1,\ldots ,N-1.
\]
\item[{\rm 2)}] 3 types of configuration are possible:
\[
x_{0}<x_{1}<\dots<x_{N}\Longleftrightarrow f<\frac{\omega_{0}^{2}L+2\omega_{1}^{2}a}{2},
\]
\[
x_{0}=x_{1}=\dots=x_{N}\Longleftrightarrow f=\frac{\omega_{0}^{2}L+2\omega_{1}^{2}a}{2},
\]
\[
x_{0}>x_{1}>\dots>x_{N}\Longleftrightarrow f>\frac{\omega_{0}^{2}L+2\omega_{1}^{2}a}{2}.
\]
\end{itemize}
\end{theorem}

Note that condition $0\leq x_{0}<x_{1}<\dots<x_{N}\leq L$ holds iff
\[
0\leq x_{0}<x_{1}<\dots<x_{N}\leq L\Longleftrightarrow\omega_{1}^{2}a-\frac{\omega_{1}^{2}L}{N}\leq f<\frac{\omega_{0}^{2}L+2\omega_{1}^{2}a}{2}.
\]

\paragraph{Force on all particles}

Here we assume that constant force $f$ acts on each particle. 

\begin{theorem}\label{th3} Assume that $x_{N}\equiv L$ and the
potential energy is
\[
U=\frac{\omega_{0}^{2}}{2}x_{0}^{2}+\frac{\omega_{1}^{2}}{2}\sum_{i=1}^{N}(x_{i}-x_{i-1}-a)^{2}-f(x_{0}+x_{1}+\ldots +x_{N-1})
\]
where $f>0.$ Then: 

1) for any given parameters stable equilibrium exists and is unique; 

2) condition $0\leq x_{0}<x_{1}<\dots<x_{N}=L$ holds iff
\[
-\frac{2\omega_{1}^{2}(L-aN)}{N(N+1)}\leq f<\frac{2\omega_{1}^{2}\left(\omega_{0}^{2}L+a\omega_{1}^{2}\right)}{N(2\omega_{1}^{2}+\omega_{0}^{2}(N-1))}
\]
and the coordinates are given by
\[
x_{0}=\frac{\omega_{1}^{2}(L+N(N+1)f/(2\omega_{1}^{2})-Na)}{\omega_{1}^{2}+\omega_{0}^{2}N}
\]
\begin{align*}
  x_{k}-x_{k-1}&=-\frac{fk}{\omega_{1}^{2}}+\frac{\omega_{0}^{2}\bigl(L+N(N+1)f / (2\omega_{1}^{2})\bigr)+a\omega_{1}^{2}}{\omega_{1}^{2}+\omega_{0}^{2}N},\\
  &\qquad\qquad\qquad\qquad\qquad\qquad  \;\,  k=1,\ldots ,N-1
\end{align*}
\end{theorem}

Note that under scaling $a=a_{0}N^{-1},\, f=f_{0}N^{-1}, \, \omega_{1}^{2}=w_{1}^{2}N$
we will have
\[
0\leq x_{0}<x_{1}<\dots<x_{N}=L\Longleftrightarrow-\frac{2w_{1}^{2}(L-a_{0})}{1+N^{-1}}\leq f_{0}<\frac{2w_{1}^{2}\left(\omega_{0}^{2}L+a_{0}w_{1}^{2}\right)}{2w_{1}^{2}+\omega_{0}^{2}(1-N^{-1})}.
\]

\begin{theorem}\label{th4}
Assume that the potential energy is
\[
U=\frac{\omega_{0}^{2}}{2}x_{0}^{2}+\frac{\omega_{0}^{2}}{2}(x_{N}-L)^{2}+\frac{\omega_{1}^{2}}{2}\sum_{i=1}^{N}(x_{i}-x_{i-1}-a)^{2}-f(x_{0}+x_{1}+\ldots +x_{N}).
\]
Then: 

1) For any given parameters stable equilibrium exists and is unique.
Coordinates are given by:
\[
x_{0}=\frac{\omega_{1}^{2}(L+N(N+1)f/(2\omega_{1}^{2})-Na)+\omega_{1}^{2}\omega_{0}^{-2}f(N+1)}{2\omega_{1}^{2}+N\omega_{0}^{2}},
\]
\begin{align*}
  x_{k}-x_{k-1}=-\frac{fk}{\omega_{1}^{2}}&+\frac{\omega_{0}^{2}(L+N(N+1)f/(2\omega_{1}^{2}))+f(N+1)+2\omega_{1}^{2}a}{2\omega_{1}^{2}+N\omega_{0}^{2}},\\
  &\qquad\qquad\qquad \qquad\qquad\qquad \qquad \qquad k=1,\ldots ,N.
\end{align*}

2) Condition $x_{0}<x_{1}<\dots<x_{N}$ holds iff
\[
|f|<\frac{2\omega_{1}^{2}\left(\omega_{0}^{2}L+2\omega_{1}^{2}a\right)}{(N-1)\left(2\omega_{1}^{2}+N\omega_{0}^{2}\right)}.
\]

3) Condition $0\leq x_{0}<x_{1}<\dots<x_{N}\leq L$ holds iff $L\geq Na$
and
\[
\frac{-2\omega_{1}^{2}\omega_{0}^{2}(L-Na)}{\left(2\omega_{1}^{2}+N\omega_{0}^{2}\right)(N+1)}\leq f\leq\frac{\omega_{1}^{2}\omega_{0}^{2}(L-Na)}{\left(2\omega_{1}^{2}+N\omega_{0}^{2}\right)(N+1)}.
\]
\end{theorem}

Under the scaling $a=a_{0}N^{-1},\, f=f_{0}N^{-1},\, \omega_{1}^{2}=w_{1}^{2}N$
we will have
\[
0\leq x_{0}<x_{1}<\dots<x_{N}\leq L\Longleftrightarrow
\]
\[
-\frac{2w_{1}^{2}\omega_{0}^{2}(L-a_{0})}{\left(2w_{1}^{2}+\omega_{0}^{2}\right)(1+N^{-1})}\leq f_{0}\leq\frac{2w_{1}^{2}\omega_{0}^{2}(L-a_{0})}{\left(2w_{1}^{2}+\omega_{0}^{2}\right)(1+N^{-1})},\quad L>a_{0}.
\]

\paragraph{Example of regular continuum system of particles}

Here we assume that the reader knows the main definitions in the paper
\cite{LNCh_regular}. Assume that we are in the situation of Theorem
\ref{th3}. We assume further on that $\omega_{0}=0,a=L/N$.
The potential energy is the sum of two terms interaction energy and
external field energy: 
\[
U=U_{int}+U_{ext}=\frac{\omega_{1}^{2}}{2}\sum_{i=1}^{N}(x_{i}-x_{i-1}-a)^{2}-f(x_{0}+x_{1}+\ldots +x_{N-1}).
\]
We want to show that the scaling limits of $U$ for the configuration
$x_{k}=ka,k=1,\ldots ,N-1$
\[
\lim_{N\to\infty}U=\lim_{N\to\infty}U_{ext}=-\frac{f_{1}L}{2},
\]
and for the equilibrium configuration
\[
\lim_{N\to\infty}U=\lim_{N\to\infty}U_{ext}+\lim_{N\to\infty}U_{int}=-\frac{f_{1}L}{2}
\]
are equal. 

By Theorem \ref{th3} the condition $f< a\omega_{1}^{2} / N = L\omega_{1}^{2} / N^{2}$
should hold. Put
\[
f=\frac{f_{1}}{N},\quad \omega_{1}^{2}=w_{1}N,
\]
where $f_{1}<Lw_{1}$. And, by the same Theorem, as $N\to\infty$,
\[
x_{0}=\frac{N(N+1)f}{2\omega_{1}^{2}}\sim\frac{f_{1}}{2w_{1}},
\]
\[
x_{k}-x_{k-1}=-\frac{fk}{\omega_{1}^{2}}+a=-\frac{fk}{\omega_{1}^{2}}+\frac{L}{N}=\frac{L}{N}-\frac{f_{1}k}{w_{1}N^{2}}+o(N^{-2}),
\]
\[
x_{k}=-\frac{fk(k+1)}{\omega_{1}^{2}}+ka+x_{0}.
\]
Potential energy of the configuration $x_{k}=ka,\; k=1,\ldots ,N-1$ is
\[
U_{0}=-fa\frac{N(N-1)}{2}\sim-\frac{f_{1}L}{2}.
\]
Potential energy of the equilibrium configuration is
\[
  U=\frac{\omega_{1}^{2}}{2}\sum_{k=1}^{N}\left(\frac{fk}{\omega_{1}^{2}}\right)^{2}-f\sum_{k=1}^{N-1}\left(-\frac{fk(k+1)}{\omega_{1}^{2}}+ka+x_{0}\right)=
  \]
\[
=\frac{3\omega_{1}^{2}}{2}\sum_{k=1}^{N}\left(\frac{fk}{\omega_{1}^{2}}\right)^{2}+\sum_{k=1}^{N}\frac{f^{2}k}{\omega_{1}^{2}}-fa\frac{N(N-1)}{2}-fNx_{0}.
\]
Then, as $N\to\infty$,
\[
\frac{3\omega_{1}^{2}}{2}\sum_{k=1}^{N}\left(\frac{fk}{\omega_{1}^{2}}\right)^{2}=\frac{3f_{1}^{2}}{2w_{1}}\frac{1}{N}\sum_{k=1}^{N}\frac{k^{2}}{N^{2}}\to\frac{3f_{1}^{2}}{2w_{1}}\int_{0}^{1}x^{2}dx=\frac{f_{1}^{2}}{2w_{1}}.
\]
It follows that
\[
U\to-\frac{f_{1}L}{2}.
\]
Now for given $N$ and given configuration $X$, denote by $U_{N}(a,b,X),$
where $0\leq a<b\leq L$, the potential energy of all particles $x_{k}$
such that $x_{k}\in(a,b)$. The potential energy of the particle $k$
is defined as
\[
U_{k}=\frac{\omega_{1}^{2}}{4}(x_{k}-x_{k-1}-a)^{2}+\frac{\omega_{1}^{2}}{4}(x_{k+1}-x_{k}-a)^{2}-fx_{k}=
\]
\[
  =\frac{\omega_{1}^{2}}{4}\left(\frac{fk}{\omega_{1}^{2}}\right)^{2}+\frac{\omega_{1}^{2}}{4}\left(\frac{f(k+1)}{\omega_{1}^{2}}\right)^{2}
  -f\Bigl(-\frac{fk(k+1)}{\omega_{1}^{2}}+ka+x_{0}\Bigr)=
\]
\[
=-\frac{f_{1}^{2}}{2w_{1}N}-\frac{f_{1}Lk}{N^{2}}+\frac{3f_{1}^{2}}{2w_{1}}\frac{k^{2}}{N^{3}}+o(N^{-3}).
\]
Then there exists
\[
u(x)=\lim_{\Delta x\to0}\frac{1}{\Delta x}\lim_{N\to\infty}(U_{int,N}(x,x+\Delta x,X)+U_{ext,N}(x,x+\Delta x,X))
\]
correspondingly for the configuration $x_{k}=ka,k=1,\ldots ,N-1$ and
for the equilibrium configuration. 

For the configuration $X=\{x_{k}=ka,k=1,\ldots ,N-1\}$ we have
\[
  U_{int,N}(x,x+\Delta x,X)=0
\]
and
\[
U_{ext,N}(x,x+\Delta x,X)=-\frac{f_{1}}{N}\sum\frac{kL}{N},
\]
where the sum is over $k$ such that $x\leq\frac{kL}{N}\leq x+\Delta x.$
So 
\[
\lim_{N\to\infty}U_{ext,N}(x,x+\Delta x,X)=-\frac{f_{1}}{L}\int_{x}^{x+\Delta x}ydy
\]
and 
\[
u(x)=-\frac{f_{1}}{L}x.
\]
 For the equilibrium configuration $X$ we have
\[
U_{int,N}(x,x+\Delta x,X)=\frac{f_{1}^{2}}{2w_{1}}\sum\frac{k^{2}}{N^{3}}=\frac{f_{1}^{2}}{2L^{3}w_{1}}\frac{L}{N}\sum\frac{k^{2}L^{2}}{N^{2}}
\]
where the sum is over $k$ such that $x\leq kL/N \leq x+\Delta x,$
because of $x_{k}-x_{k-1}\sim L/N,$ as $N\to\infty.$ So
\[
\lim_{N\to\infty}U_{int,N}(x,x+\Delta x,X)=\frac{f_{1}^{2}}{2L^{3}w_{1}}\int_{x}^{x+\Delta x}y^{2}dy
\]
and
\[
\lim_{\Delta x\to0}\frac{1}{\Delta x}\lim_{N\to\infty}U_{int,N}(x,x+\Delta x,X)=\frac{f_{1}^{2}x^{2}}{2L^{3}w_{1}}.
\]
Similarly one can show
\[
\lim_{\Delta x\to0}\frac{1}{\Delta x}\lim_{N\to\infty}U_{ext,N}(x,x+\Delta x,X)=\frac{f_{1}^{2}x^{2}}{L^{3}w_{1}}-\frac{f_{1}}{L}x-\frac{f_{1}^{2}}{2w_{1}}.
\]
Hence, for the equilibrium configuration $X$ we have
\[
u(x)=\frac{3f_{1}^{2}x^{2}}{2L^{3}w_{1}}-\frac{f_{1}}{L}x-\frac{f_{1}^{2}}{2w_{1}}.
\]
Note that in both cases the integral $\intop_{0}^{L}u(x)dx=- f_{1}L / 2.$

\paragraph{Two edges}

Here we consider graph with two edges and common vertex which we denote
$0$. On the first edge of length $L_{1}$ there are $N_{1}$ particles
with coordinates
\begin{equation}
0\leq x_{1}<\ldots <x_{N_{1}-1}<x_{N_{1}}\equiv L_{1},\label{c1}
\end{equation}
and on the second edge of length $L_{2}$ there are $N_{2}$ particles
with coordinates
\begin{equation}
0\leq y_{1}<\ldots <y_{N_{2}}\equiv L_{2}. \label{c2}
\end{equation}
It is important that coordinates $x_{N_{1}}$ and $y_{N_{2}}$ are
tightly fixed and cannot move. Potential energy is given by
\[
  U=\frac{\omega_{2}^{2}}{2}\sum_{i=1}^{N_{2}-1}\Bigl(y_{i+1}-y_{i}-\frac{a_{2}}{M}\Bigr)^{2}
  +\frac{\omega_{1}^{2}}{2}\sum_{i=1}^{N_{1}-1}\Bigl(x_{i+1}-x_{i}-\frac{a_{1}}{M}\Bigr)^{2}+\frac{\omega_{0}^{2}}{2}\Bigl(x_{1}+y_{1}-\frac{a_{0}}{M}\Bigr)^{2}
\]
where all parameters $a_{j},\omega_{j}>0$ and $b= N/ M ,\, c=N_{1} / N <1$
are of order $1$, but $M,N_{1},N_{2},N=N_{1}+N_{2}$ are sufficiently
large but of the same order.

Put
\begin{equation}
r_{0}=\frac{L_{1}+L_{2}-a_{1}cb-a_{2}\left(1-c\right)b-a_{0}M^{-1}}{M^{-1}+b\omega_{0}^{2}\left(\omega_{1}^{-2}c+\omega_{2}^{-2}(1-c)\right)}. \label{r_0}
\end{equation}

\begin{theorem}\label{th5} 
1) For fixed parameters $a_{j},\omega_{j},b,c$ there exists equilibrium
state, satisfying conditions (\ref{c1}) and (\ref{c2}), iff the
following two inequalities hold:
\[
\frac{\omega_{0}^{2}r_{0}}{\omega_{1}^{2}}+a_{1}>0,\quad \frac{\omega_{0}^{2}r_{0}}{\omega_{2}^{2}}+a_{2}>0.
\]
In this case equilibrium state is unique and is defined by
\[
x_{k+1}-x_{k}=\Bigl(\frac{\omega_{0}^{2}r_{0}}{\omega_{1}^{2}}+a_{1}\Bigr)\frac{1}{M},
\]
\[
y_{l+1}-y_{l}=\Bigl(\frac{\omega_{0}^{2}r_{0}}{\omega_{2}^{2}}+a_{2}\Bigr)\frac{1}{M},
\]
for all $k=1,\ldots ,N_{1}-1$ and $l=1,\ldots ,N_{2}-1$.

2) This equilibrium state defines also the unique minimum of $U$,
which is equal to
\[
U(b,c,M)=\frac{\omega_{0}^{2}(L_{1}+L_{2}-a_{1}cb-a_{2}\left(1-c\right)b-a_{0}M^{-1})^{2}}{2bM(\omega_{0}^{2}\left(\omega_{1}^{-2}c+\omega_{2}^{-2}(1-c)\right)+(bM)^{-1})}.
\]
 \end{theorem}

\begin{remark}
To get macroscopic (of the order $1$) values for $U$ one should
scale also all frequencies as $\omega_{i}^{2}\sim c_{i}M$. Then we
will get finite ``thermodynamic limit'' of $U$ as $M\to\infty$. 
\end{remark}

\section{Proofs}

\paragraph{Proof of Theorem \ref{th0}}

We get linear system of $N$ equations for the forces $F_{k}$ acting
on the particles $k=1,2,\ldots ,N$, using the fact that the total force
on each of the particles $k=0,1,\ldots ,N-1$ equals zero:

\begin{equation}
F_{0}=-\omega_{0}^{2}x_{0}+\omega_{1}^{2}(x_{1}-x_{0}-a)+f=0, \label{0st}
\end{equation}
\begin{equation}
F_{k}=\omega_{1}^{2}(x_{k+1}-x_{k}-a)-\omega_{1}^{2}(x_{k}-x_{k-1}-a)=0,\quad k=1,\ldots ,N-1, \label{1st}
\end{equation}
where the coordinates satisfy the conditions: $x_{k}\leq L,k=1,\ldots ,N-1$
and $x_{N}=L$.

From equation (\ref{1st}) it follows that the difference $x_{k}-x_{k-1}$
does not depend on $k=1,\ldots ,N.$ Put $r=x_{k}-x_{k-1}-a.$ Then 
\[
x_{k}=x_{0}+k(r+a),\quad k=1,\ldots ,N.
\]
Using the condition $x_{N}=L$ we get one more equation:
\[
x_{N}=x_{0}+N(r+a)=L\Longleftrightarrow r+a=\frac{L-x_{0}}{N}.
\]
Finally we get system of two linear equations for two unknowns $x_{0},r$:
\begin{equation}
-\omega_{0}^{2}x_{0}+\omega_{1}^{2}r+f=0\label{1_1st}
\end{equation}
\begin{equation}
x_{0}+N(r+a)=L. \label{1_2st}
\end{equation}
Multiplying second equation on $\omega_{0}^{2}$ and adding both equations
we can find $r$
\[
(\omega_{1}^{2}+\omega_{0}^{2}N)r+\omega_{0}^{2}Na+f=\omega_{0}^{2}L\Longleftrightarrow r=\frac{\omega_{0}^{2}L-f-\omega_{0}^{2}Na}{\omega_{1}^{2}+\omega_{0}^{2}N}.
\]
Then
\[
x_{k}-x_{k-1}=r+a=\frac{\omega_{0}^{2}L-f-\omega_{0}^{2}Na}{\omega_{1}^{2}+\omega_{0}^{2}N}+a=\frac{\omega_{0}^{2}L+a\omega_{1}^{2}-f}{\omega_{1}^{2}+\omega_{0}^{2}N}.
\]
From equation (\ref{1_1st})
\[
  -\omega_{0}^{2}x_{0}+\omega_{1}^{2}\Bigl(\frac{L-x_{0}}{N}-a\Bigr)+f
  =0\Longleftrightarrow-\Bigl(\omega_{0}^{2}+\frac{\omega_{1}^{2}}{N}\Bigr)x_{0}+\omega_{1}^{2}\Bigl(\frac{L}{N}-a\Bigr)+f=0
\]
we get
\[
x_{0}=\frac{\omega_{1}^{2}(L-Na)+Nf}{\omega_{1}^{2}+\omega_{0}^{2}N}.
\]
Thus, the solution of the system (\ref{0st}), (\ref{1st}) is
\begin{equation}
x_{k}=x_{0}+k\frac{\omega_{0}^{2}L+a\omega_{1}^{2}-f}{\omega_{1}^{2}+\omega_{0}^{2}N}=\frac{(\omega_{1}^{2}+\omega_{0}^{2}k)L-\omega_{1}^{2}(N-k)a+(N-k)f}{\omega_{1}^{2}+\omega_{0}^{2}N}\label{3st}
\end{equation}
where $k=1,\ldots ,N-1.$ 

According to (\ref{3st}) the condition that $x_{k}\leq L$ for $k=1,\ldots ,N-1$
is equivalent to the condition $f\leq\omega_{0}^{2}L+\omega_{1}^{2}a.$
For $f<\omega_{0}^{2}L+\omega_{1}^{2}a$ we get the condition
\[
x_{0}<x_{1}<\dots<x_{N-1}<x_{N}=L.
\]
For $f=\omega_{0}^{2}L+\omega_{1}^{2}a$ we get the solution $x_{k}\equiv L,k=0,1,\ldots ,N.$
For $f>\omega_{0}^{2}L+\omega_{1}^{2}a$ the solution (\ref{3st})
does not have sense as by (\ref{3st}) we have
\[
x_{0}>x_{1}>\dots>x_{N-1}>x_{N}=L,
\]
what is impossible if we assume that the particles cannot jump through
fixed particle $x_{N}=L.$ Thus, for $f>\omega_{0}^{2}L+\omega_{1}^{2}a$
we get the equilibrium as $x_{k}\equiv L,k=0,1,\ldots ,N.$

Note that the uniqueness of equilibrium configuration for $f\leq\omega_{0}^{2}L+\omega_{1}^{2}a$
follows from uniqueness of solution of the linear system (\ref{1_1st}), (\ref{1_2st}). 

As the equilibrium configuration is unique and the potential energy
$U$ is bounded from below (and unbounded from above) quadratic function,
then this point is the unique minimum of $U.$

\paragraph{Proof of Theorem \ref{th1}}

We have the following system of equations for the forces $F_{k}$ acting
on the particles $k=1,2,\ldots ,N$:
\begin{equation}
F_{0}=-\omega_{0}^{2}x_{0}+\omega_{1}^{2}(x_{1}-x_{0}-a)+f=0,\label{4th}
\end{equation}
\begin{equation}
F_{k}=\omega_{1}^{2}(x_{k+1}-x_{k}-a)-\omega_{1}^{2}(x_{k}-x_{k-1}-a)=0,\quad k=1,\ldots ,N-1\label{5th}
\end{equation}
\begin{equation}
F_{N}=-\omega_{0}^{2}(x_{N}-L)-\omega_{1}^{2}(x_{N}-x_{N-1}-a)=0.\label{6th}
\end{equation}
From (\ref{5th}) it follows that the difference $x_{k}-x_{k-1}$
does not depend on $k.$ Put $r=x_{k}-x_{k-1}-a.$ Then
\[
x_{k}=x_{0}+k(r+a),\quad k=1,\ldots ,N.
\]
Substituting $x_{N}=x_{0}+N(r+a)$ and $r=x_{N}-x_{N-1}-a$ to equation
(\ref{6th}) we get system of two equations with respect to $x_{0},r$:
\begin{equation}
-\omega_{0}^{2}x_{0}+\omega_{1}^{2}r+f=0,\label{7th}
\end{equation}
\begin{equation}
-\omega_{0}^{2}x_{0}-(\omega_{1}^{2}+N\omega_{0}^{2})r-\omega_{0}^{2}Na+\omega_{0}^{2}L=0. \label{8th}
\end{equation}
Subtracting second equation from the first we get, we find $r$:
\[
(2\omega_{1}^{2}+N\omega_{0}^{2})r+f+\omega_{0}^{2}Na-\omega_{0}^{2}L=0\Longleftrightarrow r=\frac{\omega_{0}^{2}(L-Na)-f}{2\omega_{1}^{2}+N\omega_{0}^{2}},
\]
\begin{equation}
x_{k}-x_{k-1}=r+a=\frac{\omega_{0}^{2}L+2\omega_{1}^{2}a-f}{2\omega_{1}^{2}+N\omega_{0}^{2}},\quad k=1,\ldots ,N\label{dx}
\end{equation}
and $x_{0}$ we find from equation (\ref{7th})
\[
x_{0}=\frac{\omega_{1}^{2}r+f}{\omega_{0}^{2}}=\frac{\omega_{1}^{2}\left(L-Na\right)+Nf+\omega_{1}^{2}\omega_{0}^{-2}f}{2\omega_{1}^{2}+N\omega_{0}^{2}}.
\]
Finally also for other coordinates in equilibrium
\begin{align*}
  x_{k}&=x_{0}+k\frac{\omega_{0}^{2}L+2\omega_{1}^{2}a-f}{2\omega_{1}^{2}+N\omega_{0}^{2}}\\
  &=\frac{(\omega_{1}^{2}+k\omega_{0}^{2})L+(N-k)f+\omega_{1}^{2}\omega_{0}^{-2}f-\omega_{1}^{2}(N-k)a+k\omega_{1}^{2}a}{2\omega_{1}^{2}+N\omega_{0}^{2}},
\\
x_{N}&=\frac{(\omega_{1}^{2}+N\omega_{0}^{2})L+\omega_{1}^{2}\omega_{0}^{-2}f+N\omega_{1}^{2}a}{2\omega_{1}^{2}+N\omega_{0}^{2}}.
\end{align*}
Accordingly to (\ref{dx}), we get the following equivalences 
\[
x_{0}<x_{1}<\ldots <x_{N}\Longleftrightarrow f<\omega_{0}^{2}L+2\omega_{1}^{2}a;
\]
\[
x_{0}=x_{1}=\ldots =x_{N}\Longleftrightarrow f=\omega_{0}^{2}L+2\omega_{1}^{2}a;
\]
\[
x_{0}>x_{1}>\ldots >x_{N}\Longleftrightarrow f>\omega_{0}^{2}L+2\omega_{1}^{2}a;
\]
\[
x_{N}\leq L\Longleftrightarrow\frac{(\omega_{1}^{2}+N\omega_{0}^{2})L+\omega_{1}^{2}\omega_{0}^{-2}f+N\omega_{1}^{2}a}{2\omega_{1}^{2}+N\omega_{0}^{2}}\leq L\Longleftrightarrow f\leq\omega_{0}^{2}(L-Na);
\]
\[
x_{0}\geq0\Longleftrightarrow\frac{\omega_{1}^{2}\left(L-Na\right)+Nf+\omega_{1}^{2}\omega_{0}^{-2}f}{2\omega_{1}^{2}+N\omega_{0}^{2}}\geq0\Longleftrightarrow f\geq-\frac{\omega_{1}^{2}\left(L-Na\right)}{N+\omega_{1}^{2}\omega_{0}^{-2}}.
\]
It follows that the condition, $0\leq x_{0}<x_{1}<\ldots <x_{N}\leq L$
is equivalent to
\[
-\frac{\omega_{1}^{2}\left(L-Na\right)}{N+\omega_{1}^{2}\omega_{0}^{-2}}\leq f\leq\omega_{0}^{2}(L-Na)
\]
as from $f\leq\omega_{0}^{2}(L-Na)$ it follows that $f<\omega_{0}^{2}L+2\omega_{1}^{2}a.$

As in Theorem \ref{th0} this equilibrium configuration is unique
and stable.

\paragraph{Proof of Theorem \ref{th2}}

The system of equations is
\[
F_{0}=-\omega_{0}^{2}x_{0}+\omega_{1}^{2}(x_{1}-x_{0}-a)+f=0,
\]
\[
F_{k}=\omega_{1}^{2}(x_{k+1}-x_{k}-a)-\omega_{1}^{2}(x_{k}-x_{k-1}-a)=0,\quad k=1,\ldots ,N-1,
\]
\[
F_{N}=-\omega_{0}^{2}(x_{N}-L)-\omega_{1}^{2}(x_{N}-x_{N-1}-a)-f=0.
\]
Also $x_{k}-x_{k-1}$ does not depend on $k,$ and we put 
\[
x_{k}=x_{0}+k(r+a),\quad k=1,\ldots ,N
\]
where $r=x_{k}-x_{k-1}-a.$ 

Substituting $x_{N}=x_{0}+N(r+a),r=x_{N}-x_{N-1}-a$ to equation $F_{N}=0$,
we get the system of two equations w.r.t.\ $x_{0},r$:
\[
-\omega_{0}^{2}x_{0}+\omega_{1}^{2}r+f=0,
\]
\[
-\omega_{0}^{2}x_{0}-(\omega_{1}^{2}+N\omega_{0}^{2})r-\omega_{0}^{2}Na+\omega_{0}^{2}L-f=0,
\]
from where we get
\[
r=\frac{\omega_{0}^{2}(L-Na)-2f}{2\omega_{1}^{2}+N\omega_{0}^{2}}, \quad x_{0}=\frac{\omega_{1}^{2}\left(L-Na\right)+Nf}{2\omega_{1}^{2}+N\omega_{0}^{2}}.
\]
It follows $x_{0}\geq0\Longleftrightarrow$$\omega_{1}^{2}\left(L-Na\right)+Nf>0\Longleftrightarrow$$f\geq\omega_{1}^{2}(a- L/N).$
And then
\[
x_{k}-x_{k-1}=r+a=\frac{\omega_{0}^{2}L+2\omega_{1}^{2}a-2f}{2\omega_{1}^{2}+N\omega_{0}^{2}},\quad k=1,\ldots ,N,
\]
and
\begin{align*}
  x_{k}&=x_{0}+k\frac{\omega_{0}^{2}L+2\omega_{1}^{2}a-2f}{2\omega_{1}^{2}+N\omega_{0}^{2}}=\\
  &=\frac{(\omega_{1}^{2}+k\omega_{0}^{2})L+(N-2k)f-\omega_{1}^{2}(N-k)a+k\omega_{1}^{2}a}{2\omega_{1}^{2}+N\omega_{0}^{2}}.
\end{align*}
Thus, 
\[
x_{0}<x_{1}<\ldots <x_{N}\Longleftrightarrow2f<\omega_{0}^{2}L+2\omega_{1}^{2}a.
\]
In particular,
\[
x_{N}=\frac{(\omega_{1}^{2}+N\omega_{0}^{2})L-Nf+N\omega_{1}^{2}a}{2\omega_{1}^{2}+N\omega_{0}^{2}},
\]
\[
x_{N}\leq L\Longleftrightarrow\frac{(\omega_{1}^{2}+N\omega_{0}^{2})L-Nf+N\omega_{1}^{2}a}{2\omega_{1}^{2}+N\omega_{0}^{2}}\leq L\Longleftrightarrow f\geq\omega_{1}^{2}\Bigl(a-\frac{L}{N}\Bigr)a.
\]
Finally
\[
0\leq x_{0}<x_{1}<\ldots <x_{N}\leq L\Longleftrightarrow\omega_{1}^{2}\Bigl(a-\frac{L}{N}\Bigr)a\leq f<\frac{\omega_{0}^{2}L+2\omega_{1}^{2}a}{2}.
\]

\paragraph{Proof of Theorem \ref{th3}}

The system of equations is
\begin{equation}
F_{0}=-\omega_{0}^{2}x_{0}+\omega_{1}^{2}(x_{1}-x_{0}-a)+f=0, \label{9th}
\end{equation}
\begin{equation}
F_{k}=\omega_{1}^{2}(x_{k+1}-x_{k}-a)-\omega_{1}^{2}(x_{k}-x_{k-1}-a)+f=0,\;\, k=1,\ldots ,N-1,\label{10th}
\end{equation}
where the coordinates $x_{k}$ should also satisfy the conditions:
$x_{k}\leq L,k=1,\ldots ,N-1$ and $x_{N}=L.$ 

Put $r_{k}=x_{k}-x_{k-1}-a.$ Then the system (\ref{9th}) and (\ref{10th})
can be rewritten as:
\[
r_{1}=-\frac{f}{\omega_{1}^{2}}+\frac{\omega_{0}^{2}x_{0}}{\omega_{1}^{2}}, \quad r_{k}=-\frac{f}{\omega_{1}^{2}}+r_{k-1},\quad k=2,\ldots ,N.
\]
Then
\[
r_{k}=-\frac{f}{\omega_{1}^{2}}k+\frac{\omega_{0}^{2}x_{0}}{\omega_{1}^{2}},\quad k=1,\ldots ,N.
\]
Using the last formula we write $x_{k},k=1,\ldots ,N-1$ in terms of $x_{0}:$
\[
x_{k}=x_{0}+\sum_{l=1}^{k}\left(x_{k}-x_{k-1}-a\right)+ka=x_{0}+\sum_{l=1}^{k}r_{l}+ka=
\]
\[
  =x_{0}+\sum_{l=1}^{k}\left(-\frac{f}{\omega_{1}^{2}}l+\frac{\omega_{0}^{2}x_{0}}{\omega_{1}^{2}}\right)+ka
  =x_{0}+k\Bigl(\frac{\omega_{0}^{2}x_{0}}{\omega_{1}^{2}}+a\Bigr)-\frac{f}{\omega_{1}^{2}}\frac{k(k+1)}{2}.
\]
Then,
\[
x_{k}=x_{0}+k\Bigl(\frac{\omega_{0}^{2}x_{0}}{\omega_{1}^{2}}+a\Bigr)-\frac{k(k+1)f}{2\omega_{1}^{2}},
\]
\begin{equation}
x_{k}-x_{k-1}=r_{k}+a=-\frac{f}{\omega_{1}^{2}}k+\frac{\omega_{0}^{2}x_{0}}{\omega_{1}^{2}}+a.\label{dx1}
\end{equation}
From the condition $x_{N}=L$ we get the equation for $x_{0}$
\[
x_{0}+N\Bigl(\frac{\omega_{0}^{2}x_{0}}{\omega_{1}^{2}}+a\Bigr)-\frac{N(N+1)f}{2\omega_{1}^{2}}=L\Longleftrightarrow
\]
\[
x_{0}=\frac{\omega_{1}^{2}\bigl( L+ N(N+1)f / (2\omega_{1}^{2}) -Na\bigr)}{\omega_{1}^{2}+\omega_{0}^{2}N}.
\]
It is easy to see that 
\[
x_{0}\geq0\Longleftrightarrow\omega_{1}^{2}\Bigl(L+\frac{N(N+1)f}{2\omega_{1}^{2}}-Na\Bigr)>0\Longleftrightarrow f\geq-\frac{2\omega_{1}^{2}(L-Na)}{N(N+1)}.
\]
Then by (\ref{dx1}) we have
\[
x_{k}-x_{k-1}=-\frac{fk}{\omega_{1}^{2}}+\frac{\omega_{0}^{2}\bigl(L+ N(N+1)f / (2\omega_{1}^{2})\bigr)+a\omega_{1}^{2}}{\omega_{1}^{2}+\omega_{0}^{2}N},\;k=1,\ldots ,N-1.
\]
Assume that $f>0.$ Then the condition $x_{0}<x_{1}<\ldots <x_{N-1}<L$
is equivalent to 
\[
-\frac{fk}{\omega_{1}^{2}}+\frac{\omega_{0}^{2}(L+ N(N+1)f / (2\omega_{1}^{2}))+a\omega_{1}^{2}}{\omega_{1}^{2}+\omega_{0}^{2}N}>0,\;\forall k=1,\ldots ,N\Longleftrightarrow
\]
\[
\frac{\omega_{0}^{2}(L+ N(N+1)f / (2\omega_{1}^{2}))+a\omega_{1}^{2}}{\omega_{1}^{2}+\omega_{0}^{2}N}>\frac{fN}{\omega_{1}^{2}}\Longleftrightarrow f<\frac{2\omega_{1}^{2}\left(\omega_{0}^{2}L+a\omega_{1}^{2}\right)}{N(2\omega_{1}^{2}+\omega_{0}^{2}(N-1))}.
\]
Thus for $f>0$ the condition $0\leq x_{0}<x_{1}<\ldots <x_{N-1}<x_{N}=L$
is equivalent to
\[
-\frac{2\omega_{1}^{2}(L-Na)}{N(N+1)}\leq f<\frac{2\omega_{1}^{2}\left(\omega_{0}^{2}L+a\omega_{1}^{2}\right)}{N(2\omega_{1}^{2}+\omega_{0}^{2}(N-1))}.
\]
If $f<0$ the condition $x_{k}>x_{k-1}$ for any $k=1,\ldots ,N$ is equivalent
to
\[
-\frac{fk}{\omega_{1}^{2}}+\frac{\omega_{0}^{2}(L+ N(N+1)f / (2\omega_{1}^{2}))+a\omega_{1}^{2}}{\omega_{1}^{2}+\omega_{0}^{2}N}>0\;\forall k=1,\ldots ,N\Longleftrightarrow
\]
\[
-\frac{f}{\omega_{1}^{2}}+\frac{\omega_{0}^{2}(L+ N(N+1)f / (2\omega_{1}^{2}))+a\omega_{1}^{2}}{\omega_{1}^{2}+\omega_{0}^{2}N}>0\Longleftrightarrow f<-\frac{2\omega_{1}^{2}(\omega_{0}^{2}L+a\omega_{1}^{2})}{\omega_{0}^{2}N(N-1)-2\omega_{1}^{2}}.
\]
And in general, for $f<0$ the condition $0\leq x_{0}<x_{1}<\ldots <x_{N-1}<x_{N}=L$
is equivalent to 
\[
-\frac{2\omega_{1}^{2}(L-Na)}{N(N+1)}\leq f<-\frac{2\omega_{1}^{2}(\omega_{0}^{2}L+a\omega_{1}^{2})}{\omega_{0}^{2}N(N-1)-2\omega_{1}^{2}}.
\]

\paragraph{Proof of Theorem \ref{th4}}

The equations are
\begin{equation}
F_{0}=-\omega_{0}^{2}x_{0}+\omega_{1}^{2}(x_{1}-x_{0}-a)+f=0,\label{11th}
\end{equation}
\begin{equation}
F_{k}=\omega_{1}^{2}(x_{k+1}-x_{k}-a)-\omega_{1}^{2}(x_{k}-x_{k-1}-a)+f=0,\; k=1,\ldots ,N-1,\label{12th}
\end{equation}
\begin{equation}
F_{N}=-\omega_{0}^{2}(x_{N}-L)-\omega_{1}^{2}r_{N}+f=0.\label{13th}
\end{equation}
Putting $r_{k}=x_{k}-x_{k-1}-a$ and as Theorem \ref{th3}, from equations
(\ref{11th}), (\ref{12th}) we find
\begin{equation}
r_{k}=-\frac{fk}{\omega_{1}^{2}}+\frac{\omega_{0}^{2}x_{0}}{\omega_{1}^{2}}\Longleftrightarrow x_{k}-x_{k-1}=-\frac{fk}{\omega_{1}^{2}}+\frac{\omega_{0}^{2}x_{0}}{\omega_{1}^{2}}+a, \label{dx2}
\end{equation}
\[
x_{k}=x_{0}+k\Bigl(\frac{\omega_{0}^{2}x_{0}}{\omega_{1}^{2}}+a\Bigr)-\frac{f}{\omega_{1}^{2}}\frac{k(k+1)}{2}
\]
for $k=1,\ldots ,N.$ Substituting 
\[
  x_{N}=x_{0}+N\Bigl(\frac{\omega_{0}^{2}x_{0}}{\omega_{1}^{2}}+a\Bigr)-\frac{N(N+1)f}{2\omega_{1}^{2}},\quad
  r_{N}=-\frac{f}{\omega_{1}^{2}}N+\frac{\omega_{0}^{2}x_{0}}{\omega_{1}^{2}}
\]
to (\ref{13th}), we can get $x_{0}:$
\begin{equation}
x_{0}=\frac{\omega_{1}^{2}(L+ N(N+1)f / (2\omega_{1}^{2}) -Na)+\omega_{1}^{2}\omega_{0}^{-2}f(N+1)}{2\omega_{1}^{2}+N\omega_{0}^{2}}. \label{x0}
\end{equation}
According to (\ref{dx2}) we have
\[
x_{k}-x_{k-1}=-\frac{fk}{\omega_{1}^{2}}+\frac{\omega_{0}^{2}x_{0}}{\omega_{1}^{2}}+a=
\]
\begin{equation}
  =-\frac{fk}{\omega_{1}^{2}}+\frac{\omega_{0}^{2}(L+ N(N+1)f / (2\omega_{1}^{2}))+f(N+1)+2\omega_{1}^{2}a}{2\omega_{1}^{2}+N\omega_{0}^{2}},\;k=1,\ldots ,N.
  \label{dx3}
\end{equation}
For $f>0$ the condition $x_{k}>x_{k-1}$ for any $k=1,\ldots ,N$ is
equivalent to
\[
\frac{\omega_{0}^{2}(L+ N(N+1)f / (2\omega_{1}^{2}))+f(N+1)+2\omega_{1}^{2}a}{2\omega_{1}^{2}+N\omega_{0}^{2}}>\frac{fk}{\omega_{1}^{2}}\;\forall k=1,\ldots ,N\Longleftrightarrow
\]
\[
\frac{\omega_{0}^{2}(L+ N(N+1)f / (2\omega_{1}^{2}))+f(N+1)+2\omega_{1}^{2}a}{2\omega_{1}^{2}+N\omega_{0}^{2}}>\frac{fN}{\omega_{1}^{2}}\Longleftrightarrow
\]
\[
f<\frac{2\omega_{1}^{2}\left(\omega_{0}^{2}L+2a\omega_{1}^{2}\right)}{(N-1)(2\omega_{1}^{2}+\omega_{0}^{2}N)}.
\]
For $f<0$ the condition $x_{k}>x_{k-1}$ for any $k=1,\ldots ,N$ is
equivalent to
\[
\frac{-\omega_{0}^{2}(L+ N(N+1)f / (2\omega_{1}^{2}))-f(N+1)-2\omega_{1}^{2}a}{2\omega_{1}^{2}+N\omega_{0}^{2}}<-\frac{fk}{\omega_{1}^{2}}\;\forall k=1,\ldots ,N\Longleftrightarrow
\]
\[
\frac{-\omega_{0}^{2}(L+ N(N+1)f / (2\omega_{1}^{2}))-f(N+1)-2\omega_{1}^{2}a}{2\omega_{1}^{2}+N\omega_{0}^{2}}<-\frac{f}{\omega_{1}^{2}}\Longleftrightarrow
\]
\[
-f<\frac{2\omega_{1}^{2}\left(\omega_{0}^{2}L+2a\omega_{1}^{2}\right)}{(N-1)(2\omega_{1}^{2}+\omega_{0}^{2}N)}.
\]
And then 
\[
x_{0}<x_{1}<\ldots <x_{N}\Longleftrightarrow|f|<\frac{2\omega_{1}^{2}\left(\omega_{0}^{2}L+2a\omega_{1}^{2}\right)}{(N-1)(2\omega_{1}^{2}+\omega_{0}^{2}N)}.
\]
From (\ref{x0}), (\ref{dx3}) we have
\[
x_{N}=\frac{ N(N+1)f/2+\omega_{1}^{2}\omega_{0}^{-2}f(N+1)+\omega_{1}^{2}(L+Na)+N\omega_{0}^{2}L}{2\omega_{1}^{2}+N\omega_{0}^{2}}.
\]
Then
\[
x_{N}\leq L\Longleftrightarrow f\leq\frac{\omega_{1}^{2}\omega_{0}^{2}(L-Na)}{\left(2\omega_{1}^{2}+N\omega_{0}^{2}\right)(N+1)}.
\]
By (\ref{x0})
\[
x_{0}\geq0\Longleftrightarrow f\geq\frac{-2\omega_{1}^{2}\omega_{0}^{2}(L-Na)}{\left(2\omega_{1}^{2}+N\omega_{0}^{2}\right)(N+1)}.
\]
Denote 
\begin{align*}
  A&=\frac{2\omega_{1}^{2}\left(\omega_{0}^{2}L+2a\omega_{1}^{2}\right)}{(N-1)(2\omega_{1}^{2}+\omega_{0}^{2}N)},\\[5pt]
  B&=\frac{\omega_{1}^{2}\omega_{0}^{2}(L-Na)}{\left(2\omega_{1}^{2}+N\omega_{0}^{2}\right)(N+1)}.
\end{align*}
Note that $0<2B<A$ is equivalent to 
\[
f\leq\frac{\omega_{1}^{2}\omega_{0}^{2}(L-Na)}{\left(2\omega_{1}^{2}+N\omega_{0}^{2}\right)(N+1)}<\frac{2\omega_{1}^{2}\left(\omega_{0}^{2}L+2a\omega_{1}^{2}\right)}{(N-1)(2\omega_{1}^{2}+\omega_{0}^{2}N)}.
\]
As $2B<A$, the condition $0\leq x_{0}<x_{1}<\ldots <x_{N}\leq L$ is
equivalent to the condition that $-2B\leq f\leq B$ and $B>0$.

\paragraph{Proof of Theorem \ref{th5}}

We get the following system of equations from the condition that forces
on each particle equal zero: 
\[
\omega_{1}^{2}\Bigl(x_{k+1}-x_{k}-\frac{a_{1}}{M}\Bigr)-\omega_{1}^{2}\Bigl(x_{k}-x_{k-1}-\frac{a_{1}}{M}\Bigr)=0,\; k=2,\ldots ,N_{1}-1,
\]
\[
\omega_{1}^{2}\Bigl(x_{2}-x_{1}-\frac{a_{1}}{M}\Bigr)-\omega_{0}^{2}\Bigl(x_{1}+y_{1}-\frac{a_{0}}{M}\Bigr)=0,
\]
\[
\omega_{2}^{2}\Bigl(y_{2}-y_{1}-\frac{a_{2}}{M}\Bigr)-\omega_{0}^{2}\Bigl(x_{1}+y_{1}-\frac{a_{0}}{M}\Bigr)=0,
\]
\[
\omega_{2}^{2}\Bigl(y_{i+1}-y_{i}-\frac{a_{2}}{M}\Bigr)-\omega_{2}^{2}\Bigl(y_{i}-y_{i-1}-\frac{a_{2}}{M}\Bigr)=0,\; i=2,\ldots ,N_{2}-1.
\]
Denote
\[
R_{k}=x_{k+1}-x_{k}-\frac{a_{1}}{M},\;\, k=1,\ldots ,N_{1}-1,
\]
\[
R_{0}=x_{1}+y_{1}-\frac{a_{0}}{M},
\]
\[
Q_{i}=y_{i+1}-y_{i}-\frac{a_{2}}{M},\;\, i=1,\ldots ,N_{2}-1.
\]
Then
\[
R_{1}=\ldots =R_{N_{1}-1}=\frac{\omega_{0}^{2}R_{0}}{\omega_{1}^{2}},
\]
\[
Q_{1}=\ldots =Q_{N_{2}-1}=\frac{\omega_{0}^{2}R_{0}}{\omega_{2}^{2}},
\]
and it follows
\[
  x_{k+1}-x_{k}-\frac{a_{1}}{M}=\frac{\omega_{0}^{2}R_{0}}{\omega_{1}^{2}}\Longleftrightarrow x_{k+1}=x_{1}
  +\left(\frac{\omega_{0}^{2}(x_{1}+y_{1}- a_{0} / M)}{\omega_{1}^{2}}+\frac{a_{1}}{M}\right)k,
\]
\[
  y_{i+1}-y_{i}-\frac{a_{2}}{M}=\frac{\omega_{0}^{2}R_{0}}{\omega_{2}^{2}}\Longleftrightarrow y_{i+1}=y_{1}
  +\left(\frac{\omega_{0}^{2}(x_{1}+y_{1}- a_{0} / M)}{\omega_{2}^{2}}+\frac{a_{2}}{M}\right)i.
\]
By conditions (\ref{c1}) and (\ref{c2}) it should be 
\[
x_{k+1}>x_{k}\Longleftrightarrow\frac{\omega_{0}^{2}R_{0}}{\omega_{1}^{2}}+\frac{a_{1}}{M}>0\Longleftrightarrow R_{0}>-\frac{\omega_{1}^{2}a_{1}}{\omega_{0}^{2}M},
\]
\[
y_{k+1}>y_{k}\Longleftrightarrow\frac{\omega_{0}^{2}R_{0}}{\omega_{2}^{2}}+\frac{a_{2}}{M}>0\Longleftrightarrow R_{0}>-\frac{\omega_{2}^{2}a_{2}}{\omega_{0}^{2}M}
\]
and then 
\[
x_{1}+y_{1}-\frac{a_{0}}{M}>-\min\Bigl(\frac{\omega_{1}^{2}a_{1}}{\omega_{0}^{2}M},\frac{\omega_{2}^{2}a_{2}}{\omega_{0}^{2}M}\Bigr).
\]
The potential energy is then
\begin{align*}
U&=\frac{\omega_{0}^{4}R_{0}^{2}(N_{2}-1)}{2\omega_{2}^{2}}+\frac{\omega_{0}^{2}R_{0}^{2}}{2}+\frac{\omega_{0}^{4}R_{0}^{2}(N_{1}-1)}{2\omega_{1}^{2}}=
\\[5pt]
  &=\frac{\omega_{0}^{2}}{2}\left(\omega_{0}^{2}\left(\frac{N_{2}-1}{\omega_{2}^{2}}+\frac{N_{1}-1}{\omega_{1}^{2}}\right)+1\right)
  \Bigl(x_{1}+y_{1}-\frac{a_{0}}{M}\Bigr)^{2}\\[5pt]
 &=\frac{\omega_{0}^{2}N}{2}\left(\omega_{0}^{2}\left(\omega_{1}^{-2}c+\omega_{2}^{-2}(1-c)\right)+N^{-1}\right)\\[5pt]
  &\quad {} \times \Bigl(x_{1}+y_{1}-\frac{a_{0}}{M}\Bigr)^{2}.
\end{align*}
Finally, from equations
\[
x_{N_{1}}=x_{1}+\left(\frac{\omega_{0}^{2}(x_{1}+y_{1}- a_{0} / M)}{\omega_{1}^{2}}+\frac{a_{1}}{M}\right)cN=L_{1},
\]
\[
y_{N_{2}}=y_{1}+\left(\frac{\omega_{0}^{2}(x_{1}+y_{1}- a_{0} / M)}{\omega_{2}^{2}}+\frac{a_{2}}{M}\right)\left(1-c\right)N=L_{2},
\]
we have
\[
x_{1}+y_{1}-\frac{a_{0}}{M}=\frac{L_{1}+L_{2}-a_{1}cb-a_{2}\left(1-c\right)b-a_{0}M^{-1}}{1+bM\omega_{0}^{2}\left(\omega_{1}^{-2}c+\omega_{2}^{-2}(1-c)\right)}
\]
and it should be
\[
  \frac{L_{1}+L_{2}-a_{1}cb-a_{2}\left(1-c\right)b-a_{0}M^{-1}}{1+bM\omega_{0}^{2}\left(\omega_{1}^{-2}c+\omega_{2}^{-2}(1-c)\right)}
  >-\min\Bigl(\frac{\omega_{1}^{2}a_{1}}{\omega_{0}^{2}M},\frac{\omega_{2}^{2}a_{2}}{\omega_{0}^{2}M}\Bigr),
\]
and the potential energy is easily calculated.

\section{Distribution of kinetic and potential energies}

\subsection{Necessary definitions}

Here we give well-known definitions (more general than necessary)
for better understanding the results below.

We consider systems with $N_{0}$ particles in $R^{d}$, denote $N=dN_{0}$
the number of coordinates $x_{i}=q_{i}\in R,i=1,\ldots ,N,$ velocities
$v_{i}$, masses $m_{i}>0$ and momenta $p_{i}=m_{i}v_{i},v_{i}=\dot{x}_{i},$
of these particles. The dynamics (trajectories) $x_{i}(t),0\leq t<\infty,$
is defined by the Hamiltonian $H=T+U,$ with kinetic $T$ and potential
$U$ energies
\[
T=\sum_{i=1}^{N}\frac{m_{i}v_{i}^{2}}{2},\quad U=U_{0}(x_{1},\ldots ,x_{N})+U_{ext},
\]
where 
\[
U_{ext}=U_{ext}(t,x_{1},\ldots ,x_{N})=-\sum_{i=1}^{N}f_{i}(t)x_{i}.
\]
The equations are 
\[
m_{i}\frac{d^{2}x_{i}}{dt^{2}}=-\frac{\partial U}{\partial x_{i}}=-\frac{\partial U_{0}}{\partial x_{i}}+f_{i}(t)
\]
with initial conditions $x_{i}(0),v_{i}(0)$. Here $U_{0}$ corresponds
to interaction between particles and $f_{i}(t)$ are external forces. 

Time averages of the energies are defined as the limits (if they exist)
\[
\langle T \rangle =\lim_{t\to\infty}\frac{1}{t}\int_{0}^{t}T(s)ds,\quad \langle U \rangle =\lim_{t\to\infty}\frac{1}{t}\int_{0}^{t}U(s)ds.
\]

\paragraph{General virial theorem}

It is the following equality:
\begin{equation}
\langle T \rangle =-\sum_{i=1}^{N} \langle (f_{i},r_{i})\rangle  \label{V-F}
\end{equation}
where $f_{i}$ is the force on the $i$-th coordinate, $r_{i}$ -
its coordinate vector. 

\medskip\noindent
{\it Proof}. Let 
\[
G=\sum_{i=1}^{N}(p_{i},r_{i}).
\]
Then
\[
\dot{G}_{t}=\sum_{i=1}^{N}(f_{i},r_{i})+2T.
\]
If all $p_{i}$ and $r_{i}$ stay uniformly bounded then virial theorem
follows as
\[
\frac{1}{t}\int_{0}^{t}\dot{G}dt=\frac{1}{t}(G(t)-G(0))\to 0.
\]

\paragraph{Virial theorem for quadratic potential}

For general quadratic potential energy 
\[
U(x)=\frac{1}{2}(x,Vx),
\]
where $x\in R^{N},$ $V=(v_{ij})$ is positive definite symmetric
$(N\times N)$-matrix. Then the force $F_{i}$ on particle $i$ 
\[
F_{i}=-\nabla_{x_{i}}U(x)=-\sum_{j=1}^{N}v_{ij}x_{j}.
\]
Then
\[
\sum_{i=1}^{N}F_{i}x_{i}=-\sum_{i=1}^{N}\sum_{j=1}^{N}v_{ij}x_{i}x_{j}=-(Vx,x)=-2U(x).
\]
Put
\[
G=\sum_{i=1}^{N}m_{i}v_{i}x_{i}.
\]
Then 
\begin{equation}
\dot{G}_{t}=\sum_{i=1}^{N}m_{i}v_{i}^{2}+\sum_{i=1}^{N}m_{i}\dot{v}_{i}x_{i}=2T+\sum_{i=1}^{N}F_{i}x_{i}=2T-2U\label{vir}
\end{equation}
and for the averages
\[
2\left( \langle T \rangle - \langle U \rangle \right)= \langle \dot{G}_{t} \rangle =\lim_{t\to\infty}\frac{G(t)-G(0)}{t}=0
\]
as $G(t)$ is uniformly bounded. To see this note first that the kinetic
and potential energies are positive and due to energy conservation
are uniformly bounded. Then the system stays in bounded volume.

It follows that kinetic and potential energies are equal
\begin{equation}
\langle T \rangle =\langle U \rangle . \label{U=00003DT}
\end{equation}
Now assume also time-dependent external forces. For example, a harmonic
force $f_{i}(t)=\sin\omega_{i}t$ on particle $i$. Then the potential
energy is
\[
U(x)=\frac{1}{2}(x,Vx)-\sum_{i=1}^{N}f_{i}x_{i}.
\]
Similarly to (\ref{vir}) we get 
\[
\langle \dot{G}_{t} \rangle =2\left(\langle T \rangle - \langle U \rangle \right).
\]
Let $\nu_{1},\ldots ,\nu_{N}$ be eigenvalues of $V.$ They are positive
and assume that for all $i,j$ 
\[
\omega_{i}\neq\sqrt{\nu_{j}}.
\]
Then $\langle \dot{G}_{t} \rangle =0$ due to boundedness of $G(t)$ and $\langle T \rangle = \langle U \rangle .$

More interesting is analogs of virial theorem for local parts of a
large system of particles. For example, in biological organism (or
even in social organism) one part of the system can move more intensively
(large $<T>,$ small $<U>$) and another part could be the contrary.
We consider here simple system and try to understand when could this
be. For calculations we will use explicit calculations -- direct but
cumbersome. Our example is the following.

\subsection{Simplest system under periodic boundary force}

Consider the chain of $N+1$ particles with coordinates $x_{0}\equiv0,x_{1},\ldots ,x_{N}.$
We assume that particle $0$ is fixed at $0$ and on the particle
$N$ acts periodic force $f(t)=c\sin\omega t.$ Potential energy of
the system is 
\[
U(x)=\frac{\omega_{1}^{2}}{2}\sum_{i=1}^{N}(x_{i}-x_{i-1}-a)^{2}-x_{N}c\sin\omega t
\]
and the equations are 
\begin{align*}
  \ddot{x}_{k}&=-\omega_{1}^{2}(x_{k}-x_{k-1}-a)+\omega_{1}^{2}(x_{k+1}-x_{k}-a)\\
  &=\omega_{1}^{2}(x_{k+1}-2x_{k}+x_{k-1}), \quad k=1,\ldots ,N-1,
\\
\ddot{x}_{N}&=-\omega_{1}^{2}(x_{N}-x_{N-1}-a)+c\sin\omega t
\end{align*}
with initial conditions $x_{k}(0)=ka,v_{k}(0)=0$. 

After change 
\[
q_{k}=x_{k}-ka, \;\, k=0,1,\ldots ,N
\]
the equations will be
\[
\ddot{q}_{k}=\omega_{1}^{2}(q_{k+1}-2q_{k}+q_{k-1}), \;\, k=1,\ldots ,N-1,
\]
\[
\ddot{q}_{N}=-\omega_{1}^{2}(q_{N}-q_{N-1})+c\sin\omega t.
\]
In the matrix form they can be rewritten as
\[
\ddot{q}=-Vq+c\sin\omega te_{N},\quad e_{N}=(\delta_{1,N}, \ldots ,\delta_{N,N})=(0,\ldots ,0,1),
\]
where $V$ is the following tridiagonal matrix $N\times N$
\[
V=\left(\begin{array}{cccc}
2\omega_{1}^{2} & -\omega_{1}^{2}\\
-\omega_{1}^{2} & 2\omega_{1}^{2} & -\omega_{1}^{2}\\
\ldots  & \ldots  & \ldots  & \ldots \\
 &  & -\omega_{1}^{2} & \omega_{1}^{2}
\end{array}\right)
\]
In the last row there is $\omega_{1}^{2},$ all the rest are $2\omega_{1}^{2}.$

\paragraph{Spectrum of matrix $\boldsymbol{V}$}

Denote $\lambda_{k}$ the eigenvalues of $V$ and let $\{h_{k},k=1,\ldots ,N\}$
be the corresponding eigenvectors $Vh_{k}=\lambda_{k}h_{k}$ with
coordinates $h_{k}=(h_{k}^{(j)},j=1,\ldots ,N).$ 

\begin{lemma} The eigenvalues and eigenvectors of $V$ are 
\[
\lambda_{k}=2\omega_{1}^{2}\left(\cos\left(\frac{\pi k}{N+ 1/2}\right)+1\right), \quad k=1,\ldots ,N
\]
\[
h_{k}^{(j)}=\frac{\sin\left(\frac{j(N-k+ 1 / 2)\pi}{N+ 1/2}\right)}{\sin\left(\frac{(N-k+ 1 / 2 )\pi}{N+ 1 / 2}\right)}, \quad j=1,\ldots ,N.
\]
\end{lemma}

As all eigenvalues are positive, we can denote them as $\lambda_{k}=\nu_{k}^{2},\; k=1, \ldots ,N,$
where it will be convenient to assume all $\nu_{k}$ also positive.
Denote by $g_{k}$ the normalized eigenvectors 
\[
g_{k}=\frac{h_{k}}{\sqrt{(h_{k},h_{k})}},\;\, k=1,\ldots ,N,
\]
which form an orthonormal basis.

The energy of the system then is 
\[
H(\psi(t))=U(\psi(t))+T(\psi(t))-f(t)q_{N}.
\]
where $\psi=(q_{1},\ldots ,q_{N},p_{1},\ldots ,p_{N})$ and 
\[
U(\psi(t))=\frac{1}{2}\sum_{1\leq j,l\leq N}V_{j,l}q_{j}q_{l}=\frac{1}{2}(q,Vq),\quad T(\psi(t))=\sum_{j=1}^{N}\frac{p_{j}^{2}}{2}=\frac{1}{2}(p,p),
\]
are internal potential and kinetic energy of the system. Then the
dynamics satisfies the following system of equations:
\[
\ddot{q}_{j}=-\sum_{l}V_{j,l}q_{l}+f(t)\delta_{j,N},\;\, j=1, \ldots ,N,
\]
where $\delta_{j,N}$ is the Kronecker symbol. Let us rewrite this
in Hamiltonian form:
\[
\left\{ \begin{array}{ll}
\dot{q}_{j}=p_{j},\\
\dot{p}_{j}=-\sum_{l}V_{j,l}q_{l}+f(t)\delta_{j,N},
\end{array}\right.
\]
and in vector notation: 
\begin{equation}
\dot{\psi}=A_{0}\psi+f(t)g_{N},\label{eq_main}
\end{equation}
where 
\[
A_{0}=\left(\begin{array}{cc}
0 & E\\
-V & 0
\end{array}\right)
\]
is $(2N\times2N)$-matrix, $E$ is the unit $(N\times N)$-matrix,
and 
\[
r_{N}=(0,\ldots ,0,e_{N})^{T}\in\mathbb{R}^{2N},\quad e_{N}=(\delta_{1,N}, \ldots ,\delta_{N,N}).
\]
It is well-known that the solution of (\ref{eq_main}) is:
\begin{equation}
\psi(t)=e^{A_{0}t}\psi(0)+\int_{0}^{t}e^{A_{0}(t-s)}f(s)r_{N}ds\label{eq:solution}
\end{equation}
with
\[
e^{A_{0}t}=\left(\begin{array}{cc}
\cos(\sqrt{V}t) & (\sqrt{V})^{-1}\sin(\sqrt{V}t)\\
-\sqrt{V}\sin(\sqrt{V}t) & \cos(\sqrt{V}t)
\end{array}\right),
\]
where matrix sine and cosine are defined, similar to matrix exponent
by corresponding series. Then we can write down the solution as:
\begin{align}
  q(t)&=\cos(\sqrt{V}t)q(0)+(\sqrt{V})^{-1}\sin(\sqrt{V}t)p(0)\nonumber \\
  &\quad {} +\int_{0}^{t}f(s)(\sqrt{V})^{-1}\sin(\sqrt{V}(t-s))e_{N}ds,\label{q_t}
\end{align}
\begin{equation}
p(t)=-\sqrt{V}\sin(\sqrt{V}t)q(0)+\cos(\sqrt{V}t)p(0)+\int_{0}^{t}f(s)\cos(\sqrt{V}(t-s))e_{N}ds. \label{p_t}
\end{equation}

Let us expand the vectors $e_{N},q(0),p(0)$ in the basis of eigenvectors
of $V$:
\[
e_{N}=\sum_{k=1}^{N}(g_{k},e_{N})g_{k},\quad q(0)=\sum_{k=1}^{N}(g_{k},q(0))g_{k}, \quad p(0)=\sum_{k=1}^{N}(g_{k},p(0))g_{k}.
\]
Then, as
\[
  (\sqrt{V})^{-1}g_{k}=\frac{1}{\nu_{k}}g_{k},
\]
\[
\sin(\sqrt{V}t)g_{k}=g_{k}\sin(\nu_{k}t),
\]
\[
\cos(\sqrt{V}t)g_{k}=g_{k}\cos(\nu_{k}t),
\]
we have
\begin{align}
  q(t)&=\sum_{k=1}^{N}\Biggl[(g_{k},e_{N})\int_{0}^{t}f(s)\frac{\sin(\nu_{k}(t-s))}{\nu_{k}}ds\nonumber \\
  &\quad {} +(g_{k},q(0))\cos(\nu_{k}t)+(g_{k},p(0))\frac{\sin(\nu_{k}t)}{\nu_{k}}\Biggr]g_{k},\label{q_t-1}
\\
  p(t)&=\sum_{k=1}^{N}\Biggl[(g_{k},e_{N})\int_{0}^{t}f(s)\cos(\nu_{k}(t-s))ds\nonumber \\
  &\quad {} -(g_{k},q(0))\nu_{k}\sin(\nu_{k}t)+(g_{k},p(0))\cos(\nu_{k}t)\Biggr]g_{k}. \label{p_t-1}
\end{align}
We have to find functions 
\[
\hat{q}_{k}(t)=\nu_{k}^{-1}\int_{0}^{t}f(s)\sin(\nu_{k}(t-s))ds,
\]
\[
\hat{p}_{k}(t)=\int_{0}^{t}f(s)\cos(\nu_{k}(t-s))ds.
\]
Since $f(t)=c\sin\omega t$,  for $\omega\neq\nu_{k}$ we have
\begin{equation}
\hat{q}_{k}(t)=\frac{c\omega}{\nu_{k}^{2}-\omega^{2}}\left(\frac{\sin(\omega t)}{\omega}-\frac{\sin(\nu_{k}t)}{\nu_{k}}\right),\label{qkh}
\end{equation}
\[
\hat{p}_{k}(t)=\frac{c\omega}{\nu_{k}^{2}-\omega^{2}}\left(\cos(\omega t)-\cos(\nu_{k}t)\right).
\]
Further on we consider zero initial conditions, then coordinates and
momenta of particle $j$ are:
\begin{equation}
q_{j}(t)=\sum_{k=1}^{N}(g_{k},e_{N})(g_{k},e_{j})\hat{q}_{k}(t), \label{qj}
\end{equation}
\[
p_{j}(t)=\sum_{k=1}^{N}(g_{k},e_{N})(g_{k},e_{j})\hat{p}_{k}(t).
\]

\subsection{Kinetic energy}

Now we can find kinetic energy of particle $j$:
\[
T_{j}(t)=\frac{1}{2}\sum_{k=1}^{N}\sum_{l=1}^{N}(g_{k},e_{N})(g_{k},e_{j})(g_{l},e_{N})(g_{l},e_{j})\hat{p}_{k}(t)\hat{p}_{l}(t),
\]
where
\[
\hat{p}_{k}(t)=\frac{c\omega}{\nu_{k}^{2}-\omega^{2}}\left(\cos(\omega t)-\cos(\nu_{k}t)\right),\quad k=1,\ldots ,N,
\]
\[
\nu_{k}^{2}=2\omega_{1}^{2}\Bigl(\cos\Bigl(\frac{\pi k}{N+ 1/2}\Bigr)+1\Bigr),\quad k=1,\ldots ,N,
\]
\[
(g_{k},e_{j})=\frac{h_{k}^{(j)}}{\sqrt{(h_{k},h_{k})}}=\frac{1}{\sqrt{(h_{k},h_{k})}}\frac{\sin\left(\frac{j(N-k+1/2)\pi}{N+1/2}\right)}{\sin\left(\frac{(N-k+1/2)\pi}{N+1/2}\right)},\quad k,j=1,\ldots ,N.
\]
Note also that
\[
\sin\left(\frac{j(N-k+1/2)\pi}{N+ 1/2}\right)=\sin\left(j\pi-\frac{jk\pi}{N+1/2}\right)=(-1)^{j-1}\sin\left(\frac{jk\pi}{N+1/2}\right),
\]
\begin{equation}
(g_{k},e_{j})=\frac{1}{\sqrt{(h_{k},h_{k})}}\frac{(-1)^{j-1}\sin\left(\frac{jk\pi}{N+ 1/2}\right)}{\sin\left(\frac{k\pi}{N+ 1/2}\right)}. \label{gkj}
\end{equation}
Then
\[
(h_{k},h_{k})=\sin^{-2}\left(\frac{k\pi}{N+ 1/2}\right)\sum_{j=1}^{N}\sin^{2}\left(\frac{jk\pi}{N+ 1/2}\right).
\]
Now we want to find the mean kinetic energy of the particle $j$
\[
\langle T_{j} \rangle =\frac{ \langle p_{j}^{2}(t) \rangle }{2}.
\]

\begin{theorem} 
If $\omega\neq\nu_{k}$ for any $k$ then (for zero initial conditions)
\begin{equation}
\< T_{j}\> =\frac{1}{4}\left(\sum_{k=1}^{N}(g_{k},e_{N})(g_{k},e_{j})\frac{\omega c}{\nu_{k}^{2}-\omega^{2}}\right)^{2}+\frac{1}{4}\sum_{k=1}^{N}(g_{k},e_{N})^{2}(g_{k},e_{j})^{2}\left(\frac{\omega c}{\nu_{k}^{2}-\omega^{2}}\right)^{2}\label{tj}
\end{equation}
\end{theorem}

\medskip\noindent
{\it Proof}. By (\ref{p_t-1}),
\[
p_{j}(t)=\sum_{k=1}^{N}(g_{k},e_{N})(g_{k},e_{j})\hat{p}_{k}(t),
\]
where
\[
\hat{p}_{k}(t)=A_{k}\left(\cos(\omega t)-\cos(\nu_{k}t)\right),A_{k}=\frac{\omega c}{\nu_{k}^{2}-\omega^{2}},
\]
then we can show that
\[
\< p_{j}^{2}(t)\>=\sum_{k=1}^{N}\sum_{l=1}^{N}(g_{k},e_{N})(g_{k},e_{j})(g_{l},e_{N})(g_{l},e_{j})\< \hat{p}_{k}(t)\hat{p}_{l}(t)\>,
\]
where
\[
\<\hat{p}_{k}(t)\hat{p}_{l}(t)\> =\begin{cases}
A_{k}^{2} & k=l,\\
\frac{1}{2}A_{k}A_{l} & k\neq l.
\end{cases}
\]
In fact, for $\omega\neq\nu_{k}$
\[
\<\hat{p}_{k}^{2}(t)\> =A_{k}^{2}\< (\cos(\omega t)-\cos(\nu_{k}t))^{2}\> =A_{k}^{2}\left(\<\cos^{2}(\omega t)\> +\<\cos^{2}(\nu_{k}t)\>\right)=A_{k}^{2}
\]
and for $k\neq l$
\begin{align*}
  \<\hat{p}_{k}(t)\hat{p}_{l}(t)\> &=A_{k}A_{l}\< (\cos(\omega t)-\cos(\nu_{k}t))(\cos(\omega t)-\cos(\nu_{l}t))\> \\
  &=A_{k}A_{l}\<\cos^{2}(\omega t)\> =\frac{A_{k}A_{l}}{2}.
\end{align*}
Then 
\[
\frac{\< p_{j}^{2}(t)\> }{2}=\frac{1}{2}\sum_{k=1}^{N}(g_{k},e_{N})^{2}(g_{k},e_{j})^{2}A_{k}^{2}+
\]
\[
+\frac{1}{4}\sum_{k\neq l}^{N}(g_{k},e_{N})(g_{l},e_{N})(g_{k},e_{j})(g_{l},e_{j})A_{k}A_{l}=
\]
\[
=\frac{1}{4}\sum_{k=1}^{N}(g_{k},e_{N})^{2}(g_{k},e_{j})^{2}A_{k}^{2}+
\]
\[
+\frac{1}{4}\sum_{k,l=1}^{N}(g_{k},e_{N})(g_{l},e_{N})(g_{k},e_{j})(g_{l},e_{j})A_{k}A_{l}=
\]
\[
=\frac{1}{4}\sum_{k=1}^{N}(g_{k},e_{N})^{2}(g_{k},e_{j})^{2}A_{k}^{2}+\frac{1}{4}\left(\sum_{k=1}^{N}(g_{k},e_{N})(g_{k},e_{j})A_{k}\right)^{2}.
\]

\begin{theorem}\label{kin} 
Assume that $\omega^{2}>4\omega_{1}^{2}.$ Then there exist limits
\[
\lim_{N\to\infty}\< T_{1}\> =0,\quad \lim_{N\to\infty}\< T_{N}\> =K>0.
\]
\end{theorem}

\medskip\noindent
{\it Proof}. By (\ref{tj}),
\[
\< T_{1}\> =\frac{1}{4}\left(\sum_{k=1}^{N}(g_{k},e_{N})(g_{k},e_{1})\frac{\omega c}{\nu_{k}^{2}-\omega^{2}}\right)^{2}+\frac{1}{4}\sum_{k=1}^{N}(g_{k},e_{N})^{2}(g_{k},e_{1})^{2}\left(\frac{\omega c}{\nu_{k}^{2}-\omega^{2}}\right)^{2},
\]
\[
\< T_{N}\> =\frac{1}{4}\left(\sum_{k=1}^{N}(g_{k},e_{N})^{2}\frac{\omega c}{\nu_{k}^{2}-\omega^{2}}\right)^{2}+\frac{1}{4}\sum_{k=1}^{N}(g_{k},e_{N})^{4}\left(\frac{\omega c}{\nu_{k}^{2}-\omega^{2}}\right)^{2}.
\]
According to (\ref{gkj}),
\[
(g_{k},e_{1})=\frac{1}{\sqrt{(h_{k},h_{k})}},
\]
\[
(g_{k},e_{N})=\frac{(-1)^{N-1}}{\sqrt{(h_{k},h_{k})}}\frac{\sin\left(\frac{k\pi}{1+1/(2N)}\right)}{\sin\left(\frac{k\pi}{N+1/2}\right)}.
\]
Then
\begin{align}
  \< T_{1}\> &=\frac{1}{4}\left(\sum_{k=1}^{N}\frac{1}{(h_{k},h_{k})}\frac{\sin\left(\frac{k\pi}{1+ 1/(2N)}\right)}{\sin\left(\frac{k\pi}{N+1/2}\right)}
               \frac{\omega c}{\nu_{k}^{2}-\omega^{2}}\right)^{2}\nonumber \\
             &\quad {} +\frac{1}{4}\sum_{k=1}^{N}\frac{1}{(h_{k},h_{k})^{2}}\frac{\sin^{2}\left(\frac{k\pi}{1+ 1/(2N)}\right)}{\sin^{2}\left(\frac{k\pi}{N+1/2}\right)}
               \left(\frac{\omega c}{\nu_{k}^{2}-\omega^{2}}\right)^{2}, \label{t1}
\\
  \< T_{N}\> &=\frac{1}{4}\left(\sum_{k=1}^{N}\frac{1}{(h_{k},h_{k})}\frac{\sin^{2}\left(\frac{k\pi}{1+ 1/(2N)}\right)}{\sin^{2}\left(\frac{k\pi}{N+1/2}\right)}\frac{\omega c}{\nu_{k}^{2}-\omega^{2}}\right)^{2}\nonumber \\
  &\quad {} +\frac{1}{4}\sum_{k=1}^{N}\frac{1}{(h_{k},h_{k})^{2}}\frac{\sin^{4}\left(\frac{k\pi}{1+ 1/(2N)}\right)}{\sin^{4}\left(\frac{k\pi}{N+1/2}\right)}\left(\frac{\omega c}{\nu_{k}^{2}-\omega^{2}}\right)^{2},\label{tn}
\end{align}
where
\[
\nu_{k}^{2}=2\omega_{1}^{2}\left(\cos\left(\frac{\pi k}{N+1/2}\right)+1\right),\quad k=1,\ldots ,N,
\]
\[
(h_{k},h_{k})=\sin^{-2}\left(\frac{k\pi}{N+ 1/2}\right)\sum_{j=1}^{N}\sin^{2}\left(\frac{jk\pi}{N+ 1/2}\right),\quad k=1,\ldots ,N.
\]
Note that as $N\to\infty$
\begin{align*}
  \frac{k\pi}{N}\sum_{j=1}^{N}\sin^{2}\left(\frac{jk\pi}{N+ 1/2}\right)&\sim\frac{k\pi}{N}\sum_{j=1}^{N}\sin^{2}\left(\frac{jk\pi}{N}\right)\\
  &\to\int_{0}^{k\pi}\sin^{2}udu
\\
&=k\int_{0}^{\pi}\sin^{2}udu=\frac{k\pi}{2}.
\end{align*}
It follows
\begin{equation}
\sum_{j=1}^{N}\sin^{2}\left(\frac{jk\pi}{N+ 1/2}\right)\sim\frac{N}{2}, \quad N\to\infty. \label{s}
\end{equation}
Consider the first term in (\ref{t1})
\[
I_{1}=\sum_{k=1}^{N}\frac{1}{(h_{k},h_{k})}\frac{\sin\left(\frac{k\pi}{1+ 1/(2N)}\right)}{\sin\left(\frac{k\pi}{N+1/2}\right)}\frac{\omega c}{\nu_{k}^{2}-\omega^{2}}=
\]
\[
  =\sum_{k=1}^{N}\frac{\sin\left(\frac{k\pi}{N+ 1/2}\right)}{\sum_{j=1}^{N}\sin^{2}\left(\frac{jk\pi}{N+1/2}\right)}\sin\left(\frac{k\pi}{1+1/(2N)}\right)
  \frac{\omega c}{\nu_{k}^{2}-\omega^{2}}.
\]
Since
\begin{equation}
\sin\left(\frac{k\pi}{1+ 1/ (2N)}\right)=(-1)^{k-1}\frac{k\pi}{2N}+O(N^{-3})\label{si}
\end{equation}
we have 
\[
  I_{1}\sim S=-\frac{1}{\pi}\sum_{k=1}^{N}\frac{\pi}{N}(-1)^{k}\frac{k\pi}{N}\sin\left(\frac{k\pi}{N}\right)
  \frac{\omega c}{2\omega_{1}^{2}\left(\cos\left( \pi k/ N\right)+1\right)-\omega^{2}}.
\]
Firstly, we sum up separately in even and odd $k$. That is, we can
write
\[
S=S_{1}+S_{2},
\]
where 
\[
  S_{1}=-\frac{1}{2\pi}\sum_{k=1}^{[N/2]}\frac{2\pi}{N}\frac{2k\pi}{N}\sin\left(\frac{2k\pi}{N}\right)\frac{\omega c}{2\omega_{1}^{2}
    \left(\cos\left( 2\pi k / N \right)+1\right)-\omega^{2}},
\]
\[
  S_{2}=\frac{1}{2\pi}\sum_{k=1}^{[N/2]}\frac{2\pi}{N}\frac{(2k+1)\pi}{N}\sin\left(\frac{(2k+1)\pi}{N}\right)\frac{\omega c}{2\omega_{1}^{2}
    \left(\cos\left( (2k+1)\pi / N \right)+1\right)-\omega^{2}}.
\]
As $N\to\infty$
\begin{align*}
S_{1}&\to-\frac{1}{\pi}\int_{0}^{\pi}\frac{u\sin u\,du}{2\omega_{1}^{2}\left(\cos u+1\right)-\omega^{2}},
\\[5pt]
S_{2}&\to\frac{1}{\pi}\int_{0}^{\pi}\frac{u\sin u\,du}{2\omega_{1}^{2}\left(\cos u+1\right)-\omega^{2}}.
\end{align*}
Then, $I_{1}\to0.$ Thus the first term in (\ref{t1}) tends to $0.$

Consider now the first term in (\ref{tn}) 
\[
  I_{2}=\sum_{k=1}^{N}\frac{1}{(h_{k},h_{k})}\frac{\sin^{2}\left(\frac{k\pi}{1+ 1/(2N)}\right)}{\sin^{2}\left(\frac{k\pi}{N+ 1/2}\right)}
  \frac{\omega c}{\nu_{k}^{2}-\omega^{2}}=
\]
\[
=\sum_{k=1}^{N}\frac{\sin^{2}\left(\frac{k\pi}{1+ 1/(2N)}\right)}{\sum_{j=1}^{N}\sin^{2}\left(\frac{jk\pi}{N+ 1/2}\right)}\frac{\omega c}{\nu_{k}^{2}-\omega^{2}}.
\]
According to (\ref{s}) and (\ref{si})
\[
I_{2}\sim\frac{1}{\pi}\sum_{k=1}^{N}\frac{\pi}{N}\left(\frac{k\pi}{N}\right)^{2}\frac{\omega c}{2\omega_{1}^{2}\left(\cos\left( \pi k / N \right)+1\right)-\omega^{2}}
\]
as $N\to\infty.$ Thus
\[
I_{2}\to\frac{c}{\pi}\int_{0}^{\pi}\frac{u^{2}\,du}{2\omega_{1}^{2}\left(\cos u+1\right)-\omega^{2}}.
\]
The integral is not $0,$ as the integrand has constant sign. 

It is not difficult to show that the second terms in (\ref{t1}) and
in (\ref{tn}) tend to 0. Finally, we get

\[
\< T_{N}\>\to\frac{c^{2}}{4\pi^{2}}\left(\int_{0}^{\pi}\frac{u^{2}\,du}{2\omega_{1}^{2}\left(\cos u+1\right)-\omega^{2}}\right)^{2},\quad \< T_{1}\>\to 0.
\]

\subsection{Potential energy}

We define potential energy of the particle $j$
\[
\< U_{j}\> =\frac{\omega_{1}^{2}}{4}\left(\< (q_{j}-q_{j-1})^{2}\> +\< (q_{j+1}-q_{j})^{2}\> \right)
\]
for $j=2,\dots , N-1.$ Here $1/4$ appears because we take
only half of the interaction energy of the particle $j$ with its
neighbors.

For $j=1,N$ we have: 
\[
\< U_{1}\> =\frac{\omega_{1}^{2}}{2}\< q_{1}^{2}\> +\frac{\omega_{1}^{2}}{4}\< (q_{2}-q_{1})^{2}\>,
\]
\[
\< U_{N}\> =\frac{\omega_{1}^{2}}{4}\< (q_{N}-q_{N-1})^{2}\> -c\< q_{N}\sin\omega t\> .
\]

\begin{theorem} \label{pot}
Assume that $\omega^{2}>4\omega_{1}^{2}.$ Then the following limits
exist 
\[
\lim_{N\to\infty}\< U_{1}\> =0,
\]
\begin{align*}
  \lim_{N\to\infty}\< U_{N}\> &=\frac{1}{2}\left(\frac{\omega_{1}}{\pi}\int_{0}^{\pi}\frac{cu^{2}\,du}{2\omega_{1}^{2}\left(\cos u+1\right)-\omega^{2}}\right)^{2}\\
  &\quad {} -\frac{1}{4\pi}\int_{0}^{\pi}\frac{cu^{2}\,du}{2\omega_{1}^{2}\left(\cos u+1\right)-\omega^{2}}>0.
\end{align*}
\end{theorem}

\medskip\noindent
{\it Proof}. Using 
\[
q_{j}(t)=\sum_{k=1}^{N}(g_{k},e_{N})(g_{k},e_{j})\hat{q}_{k}(t)
\]
where
\[
\hat{q}_{k}(t)=\frac{c\omega}{\nu_{k}^{2}-\omega^{2}}\left(\frac{\sin(\omega t)}{\omega}-\frac{\sin(\nu_{k}t)}{\nu_{k}}\right)
\]
we get
\[
\< U_{1}\> =\frac{\omega_{1}^{2}}{2}\sum_{k,l=1}^{N}(g_{k},e_{N})(g_{l},e_{N})(g_{k},e_{1})(g_{l},e_{1})\<\hat{q}_{k}(t)\hat{q}_{l}(t)\> +
\]
\[
+\frac{\omega_{1}^{2}}{4}\sum_{k,l=1}^{N}(g_{k},e_{N})(g_{l},e_{N})((g_{k},e_{2})-(g_{k},e_{1}))((g_{l},e_{2})-(g_{l},e_{1}))\<\hat{q}_{k}(t)\hat{q}_{l}(t)\>,
\]
where
\[
\<\hat{q}_{k}(t)\hat{q}_{l}(t)\> =\begin{cases}
\displaystyle \frac{A_{k}^{2}}{2}\left(\omega^{-2}+\nu_{k}^{-2}\right) & k=l,\\[6pt]
\displaystyle \frac{A_{k}A_{l}}{2\omega^{2}} & k\neq l.
\end{cases}
\]
In fact, in more details
\begin{align*}
  \<\hat{q}_{k}^{2}(t)\>&=A_{k}^{2}\Bigl\<\Bigl(\frac{\sin(\omega t)}{\omega}-\frac{\sin(\nu_{k}t)}{\nu_{k}}\Bigr)^{2}\Bigr\>\\
  &=A_{k}^{2}\Bigl(\frac{1}{\omega^{2}}\<\sin^{2}(\omega t)\> +\frac{1}{\nu_{k}^{2}}\<\sin^{2}(\nu_{k}t)\>\Bigr)
\\
&=\frac{A_{k}^{2}}{2}\left(\omega^{-2}+\nu_{k}^{-2}\right)
\intertext{and for $k\neq l$}
  \<\hat{q}_{k}(t)\hat{q}_{l}(t)\> &=A_{k}A_{l}\Bigl\<\Bigl(\frac{\sin(\omega t)}{\omega}-\frac{\sin(\nu_{k}t)}{\nu_{k}}\Bigr)
                                     \Bigl(\frac{\sin(\omega t)}{\omega}-\frac{\sin(\nu_{l}t)}{\nu_{l}}\Bigr) \Bigr \>\\
  &=\frac{A_{k}A_{l}}{\omega^{2}}\<\sin^{2}(\omega t)\> =\frac{A_{k}A_{l}}{2\omega^{2}}.
\end{align*}
Finally we get
\[
\< U_{1}\> =\frac{\omega_{1}^{2}}{4\omega^{2}}\sum_{k,l=1}^{N}(g_{k},e_{N})(g_{l},e_{N})(g_{k},e_{1})(g_{l},e_{1})A_{k}A_{l}+
\]
\[
+\frac{\omega_{1}^{2}}{4}\sum_{k,l=1}^{N}(g_{k},e_{N})^{2}(g_{k},e_{1})^{2}\frac{A_{k}^{2}}{\nu_{k}^{2}}+
\]
\[
+\frac{\omega_{1}^{2}}{8\omega^{2}}\sum_{k,l=1}^{N}(g_{k},e_{N})(g_{l},e_{N})((g_{k},e_{2})-(g_{k},e_{1}))((g_{l},e_{2})-(g_{l},e_{2}))A_{k}A_{l}+
\]
\[
+\frac{\omega_{1}^{2}}{8}\sum_{k=1}^{N}(g_{k},e_{N})^{2}((g_{k},e_{2})-(g_{k},e_{1}))^{2}\frac{A_{k}^{2}}{\nu_{k}^{2}}=
\]
\[
=\frac{\omega_{1}^{2}}{4\omega^{2}}\left(\sum_{k=1}^{N}(g_{k},e_{N})(g_{k},e_{1})A_{k}\right)^{2}+\frac{\omega_{1}^{2}}{4}\sum_{k,l=1}^{N}(g_{k},e_{N})^{2}(g_{k},e_{1})^{2}\frac{A_{k}^{2}}{\nu_{k}^{2}}+
\]
\[
+\frac{\omega_{1}^{2}}{8\omega^{2}}\left(\sum_{k=1}^{N}(g_{k},e_{N})((g_{k},e_{2})-(g_{k},e_{1}))A_{k}\right)^{2}+
\]
\[
+\frac{\omega_{1}^{2}}{8}\sum_{k,l=1}^{N}(g_{k},e_{N})^{2}((g_{k},e_{2})-(g_{k},e_{1}))^{2}\frac{A_{k}^{2}}{\nu_{k}^{2}}.
\]
Using formula (\ref{gkj}) for $(g_{k},e_{j}),$ we get
\[
\< U_{1}\> =\frac{\omega_{1}^{2}}{4\omega^{2}}\left(\sum_{k=1}^{N}\frac{1}{(h_{k},h_{k})}\frac{\sin\left(\frac{Nk\pi}{N+ 1/2}\right)}{\sin\left(\frac{k\pi}{N+ 1/2}\right)}\frac{c\omega}{\nu_{k}^{2}-\omega^{2}}\right)^{2}+
\]
\[
+\frac{\omega_{1}^{2}}{4}\sum_{k=1}^{N}\frac{1}{(h_{k},h_{k})^{2}}\frac{\sin^{2}\left(\frac{Nk\pi}{N+ 1/2}\right)}{\sin^{2}\left(\frac{k\pi}{N+ 1/2}\right)}\left(\frac{c\omega}{\nu_{k}\left(\nu_{k}^{2}-\omega^{2}\right)}\right)^{2}+
\]
\[
+\frac{\omega_{1}^{2}}{8\omega^{2}}\biggl(\sum_{k=1}^{N}\frac{1}{(h_{k},h_{k})}\frac{\sin\left(\frac{Nk\pi}{N+1/2}\right)\left(-\sin\left(\frac{2k\pi}{N+1/2}\right)-\sin\left(\frac{k\pi}{N+ 1/2}\right)\right)}{\sin^{2}\left(\frac{k\pi}{N+ 1/2}\right)}\frac{c\omega}{\nu_{k}^{2}-\omega^{2}}\biggr)^{2}+
\]
\[
  +\frac{\omega_{1}^{2}}{8}\sum_{k=1}^{N}\frac{1}{(h_{k},h_{k})^{2}}\frac{\sin\left(\frac{Nk\pi}{N+ 1/2}\right)^{2}\left(\sin\left(\frac{2k\pi}{N+ 1/2}\right)+\sin\left(\frac{k\pi}{N+ 1/2}\right)\right)^{2}}{\sin^{4}\left(\frac{k\pi}{N+ 1/2}\right)} \times
\]
\[
{} \times   \left(\frac{c\omega}{\nu_{k}\left(\nu_{k}^{2}-\omega^{2}\right)}\right)^{2}.
\]
Now we can prove that $\< U_{1}\>\to 0$ as $N\to\infty,$ similarly to
the proof of the fact that $\< T_{1}\>\to 0$ in the Theorem \ref{kin}. 

Now we find 
\[
\< (q_{N}(t)-q_{N-1}(t))^{2}\> =\left(\frac{1}{2\omega^{2}}\sum_{k=1}^{N}(g_{k},e_{N})((g_{k},e_{N})-(g_{k},e_{N-1}))\frac{c\omega}{\nu_{k}^{2}-\omega^{2}}\right)^{2}+
\]
\[
+\frac{1}{2}\sum_{k=1}^{N}(g_{k},e_{N})^{2}((g_{k},e_{N})-(g_{k},e_{N-1}))^{2}\left(\frac{c\omega}{\nu_{k}\left(\nu_{k}^{2}-\omega^{2}\right)}\right)^{2},
\]
and using (\ref{gkj}), we find
\[
\< (q_{N}(t)-q_{N-1}(t))^{2}\> =
\]
\[
  =\frac{1}{2\omega^{2}}\Biggl(\sum_{k=1}^{N}\frac{1}{(h_{k},h_{k})}
  \frac{
    \sin\left(\frac{k\pi}{1+ 1/(2N)}\right)
  }
  {\sin^{2}\left(\frac{k\pi}{N+ 1/2}\right)}\times 
  \]
  \[
 {} \times    \left(\sin\left(\frac{k\pi}{1+ 1/ (2N)}\right)+\sin\left(\frac{k\pi}{1+ 3/(2(N-1))}\right)\right)
  \frac{c\omega}{\nu_{k}^{2}-\omega^{2}}\Biggr)^{2}+
\]
\[
  +\frac{1}{2}\sum_{k=1}^{N}
  \frac{1}{(h_{k},h_{k})^{2}}
  \frac{\sin^{2}\left(\frac{k\pi}{1+ 1/(2N)}\right)
    \left(\sin\left(\frac{k\pi}{1+ 1/(2N)}\right)+\sin\left(\frac{k\pi}{1+ 3/(2(N-1))}\right)\right)^{2}
  }
  {\sin^{4}\left(\frac{k\pi}{N+ 1/2}\right)} \times 
\]
\[
{} \times   \left(\frac{c\omega}{\nu_{k}\left(\nu_{k}^{2}-\omega^{2}\right)}\right)^{2}.
\]
Denote by $J_{1}$ and $J_{2}$ correspondingly the first and second
terms in the last expressions. 

Using formulas (\ref{s}), (\ref{si}) we get that as $N\to\infty$,
\[
J_{1}\sim\frac{2}{\omega^{2}\pi^{2}}\left(\sum_{k=1}^{N}\frac{\pi}{N}\left( k\pi / N \right)^{2}\frac{c\omega}{2\omega_{1}^{2}\left(\cos\left( \pi k/N \right)+1\right)-\omega^{2}}\right)^{2}\to
\]
\[
\to2\left(\frac{1}{\pi}\int_{0}^{\pi}\frac{cu^{2}\,du}{2\omega_{1}^{2}\left(\cos u+1\right)-\omega^{2}}\right)^{2}.
\]
Similarly to Theorem \ref{kin} one can show that $J_{2}\to0.$ 

We should find also the mean value
\[
\< q_{N}\sin\omega t\> =\sum_{k=1}^{N}(g_{k},e_{N})^{2}\<\hat{q}_{k}(t)\sin\omega t\> =\frac{1}{2}\sum_{k=1}^{N}(g_{k},e_{N})^{2}\frac{c}{\nu_{k}^{2}-\omega^{2}}=
\]
\[
=\frac{1}{2}\sum_{k=1}^{N}\frac{1}{(h_{k},h_{k})}\frac{\sin^{2}\left(\frac{k\pi}{1+ 1/(2N)}\right)}{\sin^{2}\left(\frac{k\pi}{N+ 1/2}\right)}\frac{c}{\nu_{k}^{2}-\omega^{2}}.
\]
Then quite similarly as for $J_{1}$ we get that, as $N\to\infty,$
\begin{align*}
  \< q_{N}\sin\omega t\>&\sim\frac{1}{2\pi}\sum_{k=1}^{N}\frac{2\pi}{N}\left(\frac{k\pi}{2N}\right)^{2}
                          \frac{c}{2\omega_{1}^{2}\left(\cos\left( \pi k/N \right)+1\right)-\omega^{2}}\\
  &\to\frac{c}{4\pi}\int_{0}^{\pi}\frac{u^{2}\,du}{2\omega_{1}^{2}\left(\cos u+1\right)-\omega^{2}}
\end{align*}
and finally
\[
\< U_{N}\>\to\frac{1}{2}\left(\frac{\omega_{1}}{\pi}\int_{0}^{\pi}\frac{cu^{2}\,du}{2\omega_{1}^{2}\left(\cos u+1\right)-\omega^{2}}\right)^{2}-\frac{1}{4\pi}\int_{0}^{\pi}\frac{cu^{2}\,du}{2\omega_{1}^{2}\left(\cos u+1\right)-\omega^{2}}
\]
as $N\to\infty.$

\subsection{Conservation of initial order of particles (no collisions)}

Assume that initial conditions are $q_{j}(0)=0\Longleftrightarrow x_{j}(0)=ja,v_{j}(0)=0.$ 

If $\omega^{2}>4\omega_{1}^{2}$ (in particular, no resonance), then
we will show that, if the constant $c$ is sufficiently small with
respect to $a$, for any $t$ the initial order will not change, that
is there will not be collisions of particles, that is
\[
x_{1}(t)<x_{2}(t)<\ldots <x_{N}(t).
\]
By (\ref{qkh}) 
\[
|\hat{q}_{k}(t)|\leq\frac{c\omega}{\omega^{2}-\nu_{k}^{2}}\left(\frac{1}{\omega}+\frac{1}{\nu_{k}}\right)=cb_{k}.
\]
As $\omega^{2}>4\omega_{1}^{2}$ the constants $b_{k}>0,$ as $\nu_{k}^{2}<4\omega_{1}^{2}.$
Then using (\ref{qj}) we get
\[
|q_{j}(t)|\leq c\sum_{k=1}^{N}|(g_{k},e_{N})(g_{k},e_{j})|b_{k}.
\]
By (\ref{gkj}) 
\begin{align*}
  |(g_{k},e_{N})(g_{k},e_{j})|&=\frac{1}{(h_{k},h_{k})}\frac{\left|\sin\left(\frac{Nk\pi}{N+ 1/2}\right)
                                \sin\left(\frac{jk\pi}{N+1/2}\right)\right|}{\sin^{2}\left(\frac{k\pi}{N+1/2}\right)}\\
  &\leq\frac{1}{(h_{k},h_{k})}\frac{\left|\sin\left(\frac{k\pi}{1+ 1/(2N)}\right)\right|}{\sin^{2}\left(\frac{k\pi}{N+ 1/2}\right)}.
\end{align*}
Further, using (\ref{s}), (\ref{si}), we get
\begin{equation}
|q_{j}(t)|\leq\frac{cB}{N^{2}}\sum_{k=1}^{N}b_{k}=\frac{cB}{N^{2}}\sum_{k=1}^{N}\left(\frac{1}{\omega^{2}-\nu_{k}^{2}}+\frac{1}{\nu_{k}}\right)\label{est}
\end{equation}
for some constant $B>0.$ Since $\omega^{2}-\nu_{k}^{2}>\omega^{2}-4\omega_{1}^{2}>0,$ we have
\[
\frac{1}{N^{2}}\sum_{k=1}^{N}\frac{1}{\omega^{2}-\nu_{k}^{2}}\leq\frac{1}{N}\frac{1}{\omega^{2}-4\omega_{1}^{2}}\to 0,\quad N\to\infty .
\]
Finally, from $\nu_{1}>\nu_{2}>\ldots >\nu_{N}$, where 
\[
\nu_{k}^{2}=2\omega_{1}^{2}\left(\cos\left(\frac{\pi k}{N+ 1/2}\right)+1\right),\quad k=1,\ldots ,N,
\]
it follows
\[
\sum_{k=1}^{N}\frac{1}{\nu_{k}}\leq\frac{N}{\nu_{N}}.
\]
Note that 
\[
\cos\left(\frac{\pi N}{N+ 1/2}\right)+1=O(N^{-2}).
\]
Hence, $\nu_{N}^{-1}=O(N)$ and
\[
\sum_{k=1}^{N}\frac{1}{\nu_{k}}=O(N^{2}).
\]
So the right hand side of the inequality (\ref{est}) is equal to
$O(1)$ as $N\to\infty.$ 

It follows that one can choose parameters $c$ and $a$ so that
for all $N$ 
\[
|q_{j}(t)|<\frac{a}{2},\quad j=1,\ldots ,N.
\]

Another case was considered in \cite{LM_Chaplygin}, where frequencies
and constant $a$ are scaled so that this property (called regularity
in \cite{LNCh_regular}) holds for all $N$. 

\section{Conclusion}

In all examples of ground states above, it can be easily proved that
for the system of equations
\begin{equation}
\frac{d^{2}x_{k}(t)}{dt^{2}}=F_{k}-\alpha_{k}v_{k}(t),\quad k=0,1,\ldots ,N,\label{dynamics}
\end{equation}
where $v_{k}=dx_{k}/dt$ and $\alpha_{k}\geq\alpha$ for some
$\alpha>0$, the following statement holds: for any initial conditions
$x_{k}(0),v_{k}(0)$ the solution converges to the corresponding minimum
of potential energy. The proof is exactly the same as in (\cite{Mal_Interval})
for Coulomb systems. More difficult is the question whether it is
true when some $\alpha_{k}=0$.

One of the next problems is the following. Assume that we define the
system to be ``healthy'' if the configuration is close to the ground
state in $l_{1}$-metrics that is if for some $\epsilon>0$ and all
$k$ we have $|x_{k}(t)-a|<\epsilon$. The question is the following:
is this domain invariant w.r.t.\ dynamics (\ref{dynamics}), or some
differences $|x_{k}(t)-x_{k-1}(t)|$ between nearest neighbors can
become too small (or even collide) or can become ``too big''. Obviously,
it is for the scaled parameters for which the ground state satisfies
condition $x_{0}<x_{1}<\ldots <x_{N}$.

\end{document}